\documentclass{emulateapj}

\usepackage{epsfig}
\newcommand{\co}{\mbox{\rm $^{12}$CO}}

\newcommand{\jone}{\mbox{($1\rightarrow0$)}}

\newcommand{\kmpers}{\mbox{km~s$^{-1}$}}

\shorttitle{Bias-free GMC Properties}
\shortauthors{Rosolowsky \& Leroy}
\begin{document}
\title{Bias-free Measurement of Giant Molecular Cloud Properties}
\author{Erik Rosolowsky\altaffilmark{1}}
\affil{Harvard-Smithsonian Center for Astrophysics, 60 Garden Street, MS-66, Cambridge, MA 02138}
\email{erosolow@cfa.harvard.edu}
\and
\author{Adam Leroy}
\affil{Department of Astronomy, 601 Campbell Hall, University of
       California at Berkeley, CA  94720}
\email{aleroy@astro.berkeley.edu}

\altaffiltext{1}{National Science Foundation Astronomy \& Astrophysics Postdoctoral Fellow}

\begin{abstract}
We review methods for measuring the sizes, line widths, and
luminosities of giant molecular clouds (GMCs) in molecular-line data
cubes with low resolution and sensitivity.  We find that moment
methods are robust and sensitive --- making full use of both position
and intensity information --- and we recommend a standard method to
measure the position angle, major and minor axis sizes, line width,
and luminosity using moment methods. Without corrections for the
effects of beam convolution and sensitivity to GMC properties, the
resulting properties may be severely biased. This is particularly true
for extragalactic observations, where resolution and sensitivity
effects often bias measured values by 40\% or more. We correct for
finite spatial and spectral resolutions with a simple deconvolution
and we correct for sensitivity biases by extrapolating properties of a
GMC to those we would expect to measure with perfect sensitivity
(i.e.~the 0~K isosurface).  The resulting method recovers the
properties of a GMC to within 10\% over a large range of resolutions
and sensitivities, provided the clouds are marginally resolved with a
peak signal-to-noise ratio greater than 10.  We note that
interferometers systematically underestimate cloud properties,
particularly the flux from a cloud.  The degree of bias depends on the
sensitivity of the observations and the $(u,v)$ coverage of the
observations.  In the Appendix to the paper we present a conservative,
new decomposition algorithm for identifying GMCs in molecular-line
observations. This algorithm treats the data in physical rather than
observational units (i.e.~parsecs rather than beams or arcseconds),
does not produce spurious clouds in the presence of noise, and is
sensitive to a range of morphologies. As a result, the output of this
decomposition should be directly comparable among disparate data sets.
\end{abstract}

\section{Introduction}

Over the last 15 years, it has become possible to observe molecular
emission in nearby galaxies with sufficient resolution and sensitivity
to distinguish individual giant molecular clouds (GMCs). The immediate
goal of such studies is to determine whether (and how) the GMCs in
other galaxies differ from those seen in the Solar neighborhood. The
most common method used to address this question has been to use
molecular-line tracers of H$_2$, in particular $^{12}$CO($1\to 0$), to
compare the macroscopic properties (size, line width, and luminosity)
of GMCs in other galaxies to those of Milky Way GMCs. Unfortunately, a
wide variety of methods have been used to reduce data from spectral
line data cubes into macroscopic GMC properties. As a result, many of
the differences between GMC populations found in the literature can be
attributed, at least partially, to observational artifacts or
methodological differences. It is therefore difficult to assess what
the real differences between GMC populations are based on the reported
data in the literature.

\begin{figure}
\begin{center}
\plotone{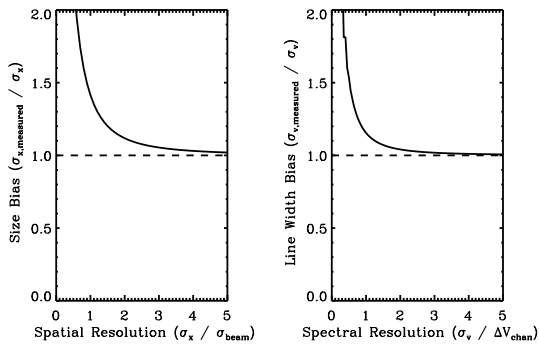} \figcaption{\label{RESBIAS} The resolution
bias for a model cloud. These two panels show the variations in the
measured spatial size (left) and line width (right) as a function of
the resolution of a data set. These plots show measurements of
Gaussian clouds convolved with a Gaussian beam and integrated across
square velocity channels. Marginally resolved data yields measurements
affected by a significant {\em resolution bias}. The spatial size bias
is encountered frequently in extragalactic GMC measurements.}
\end{center}
\end{figure}

\begin{figure*}
\begin{center}
\plotone{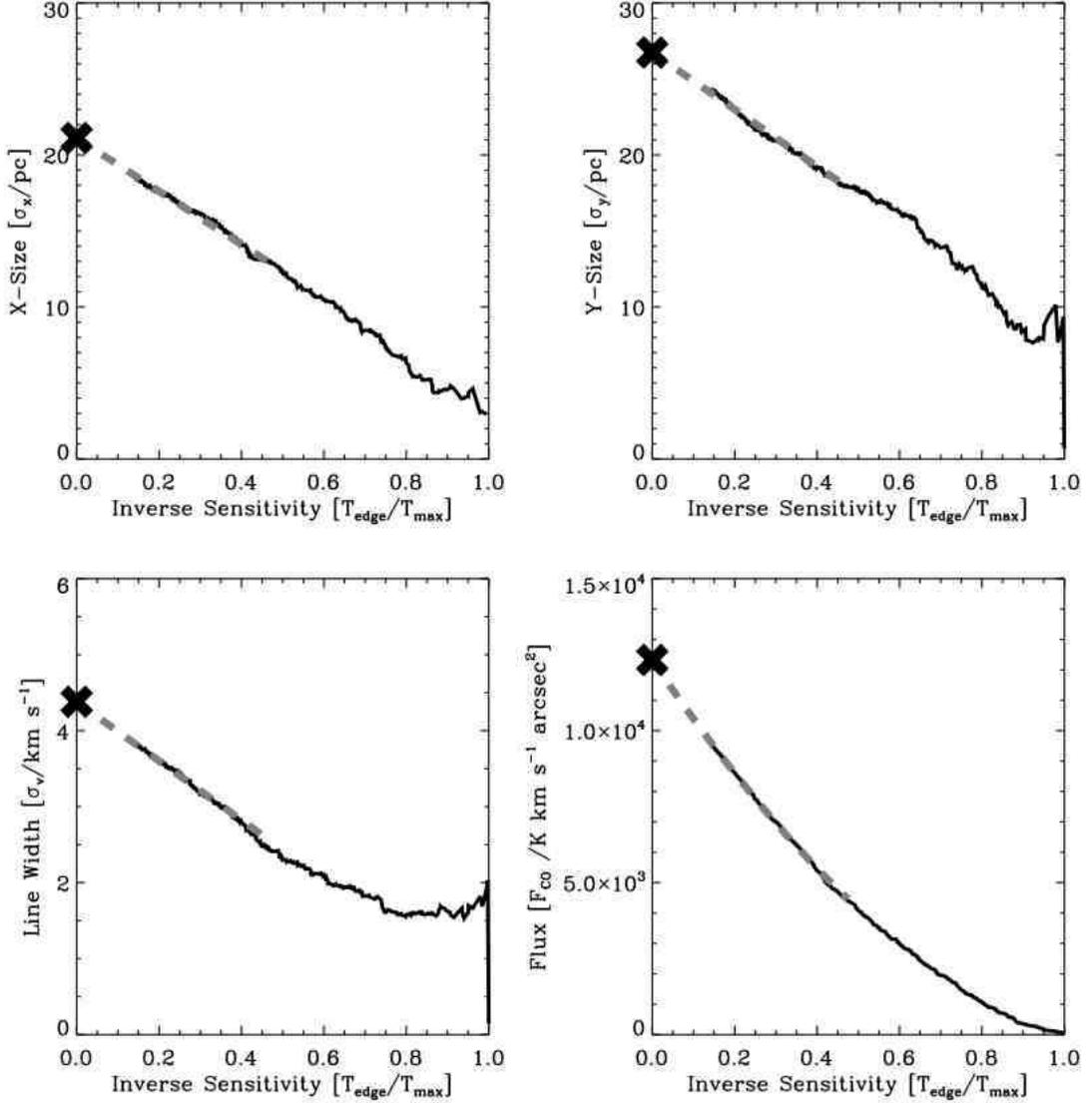}
\figcaption{\label{EXTRAPMOMS} The sensitivity bias for a bright GMC
in M~33 \citep[EPRB1 from][]{ros03}. These four panels show the
variations in the measured spatial size, line width, and flux of the
GMC for a range of sensitivities (signal-to-noise ratios). We simulate
the effect of changing sensitivity by varying the boundary isosurface,
$T_{edge}$, and measuring properties only from data within this
isosurface. All four properties are a strong function of $T_{edge}$,
so the data display a substantial {\em sensitivity bias}. The gray,
dashed lines show the extrapolation to the 0 Kelvin isosurface using a
weighted, least-squares fit to each property as a function of
$T_{edge}$.  The extrapolated value is shown as an $\times$.}
\end{center}
\end{figure*}

For GMCs that are either marginally resolved or marginally detected,
observational biases can be severe.  Figure \ref{RESBIAS} shows the
variation of the measured spatial size and line width with the
resolution for a model cloud.  Typical Galactic GMCs have sizes of a
few 10s of parsecs, comparable to the spatial resolution of many data
sets used to study extragalactic GMCs
\citep[e.g.~][]{vog87,ws90,wr91,is93,wil93,fu99,she00,ros03}.  Figure
\ref{RESBIAS} shows that when the size of the beam is comparable to
the size of the object, the measured size is much higher than the true
size of the object. Millimeter spectrometers and correlators often
have excellent frequency resolution, so the spectral resolution bias
is usually less important for GMC studies, but it can become
substantial when data are binned to increase signal-to-noise.  Figure
\ref{EXTRAPMOMS} shows that the measured spatial size, line width, and
flux of a real GMC in M~33 \citep[EPRB1 from][]{ros03} are all strong
functions of the sensitivity of the data.  We discuss another major
source of bias, the method by which emission is decomposed into GMCs,
in the Appendix.

\begin{figure*}
\begin{center}
\plottwo{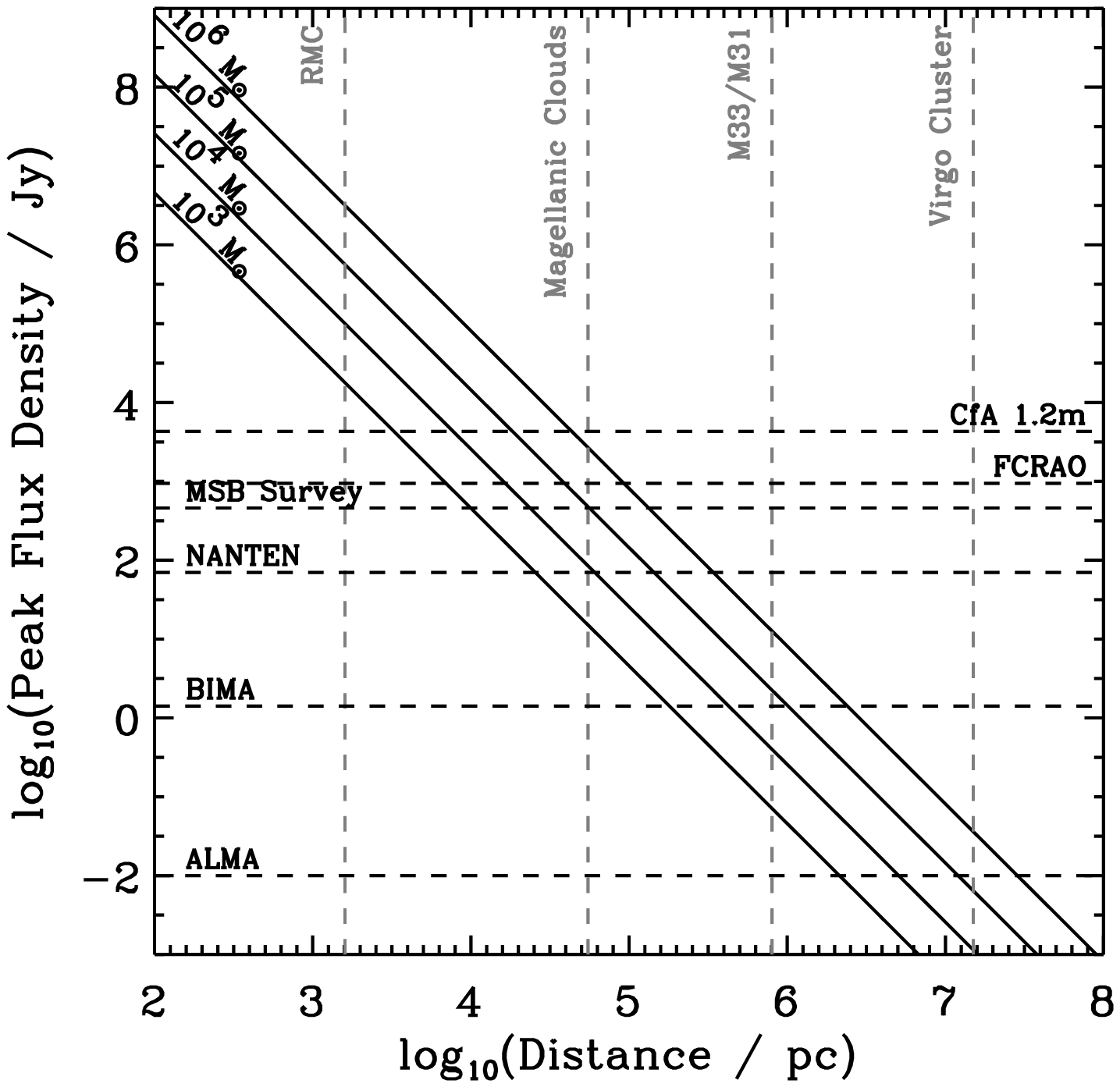}{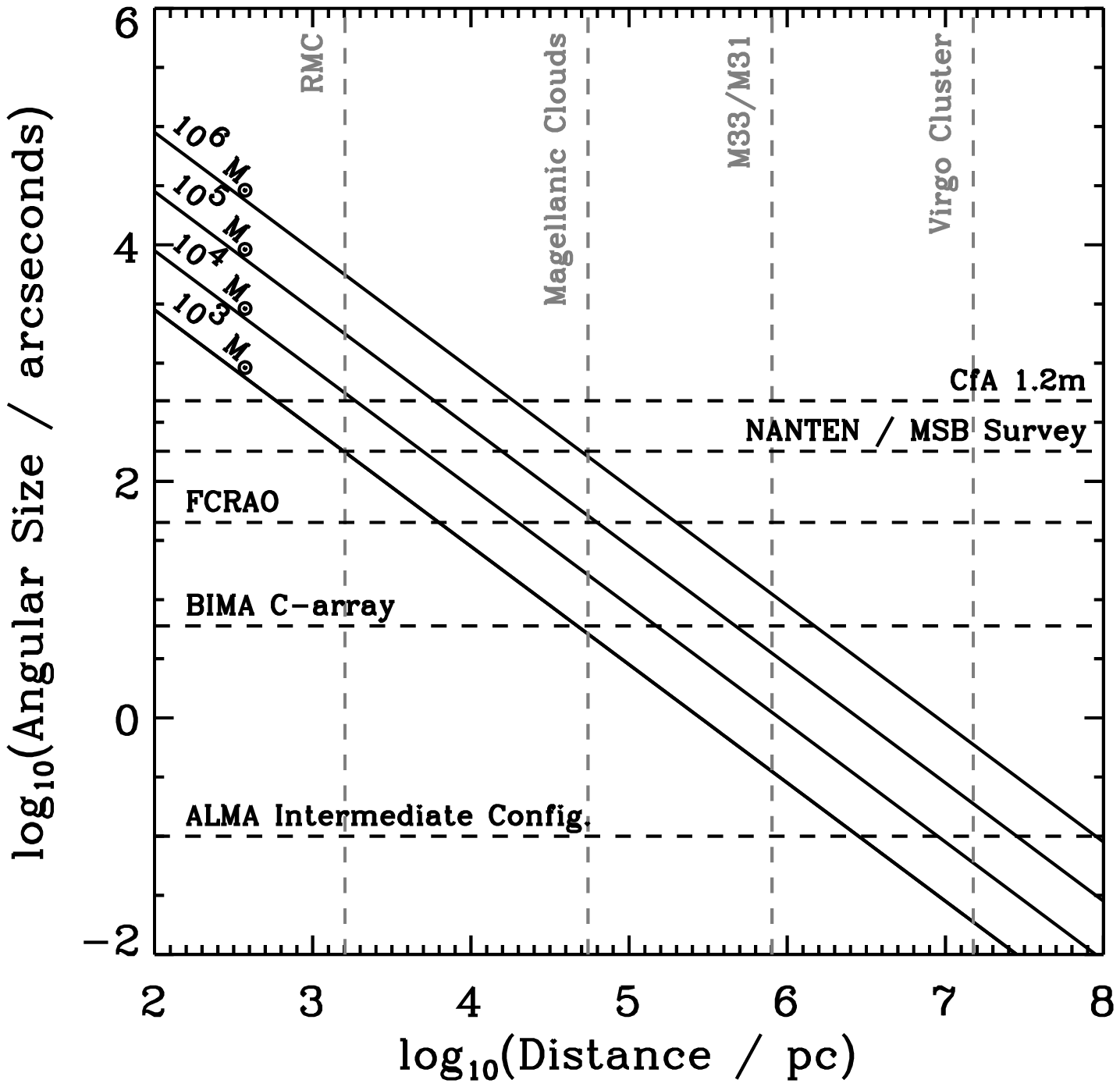} \figcaption{\label{DISTCOMP} Peak flux
density (left) and angular size (right) of molecular clouds as a
function of distance.  Diagonal lines indicate the peak flux density
and size of clouds with $10^3,10^4,10^5$ and $10^6~M_{\odot}$ (bottom
to top) from the relationships observed in the inner Milky Way
\citep{srby87}.  For reference, the diameters of these clouds are roughly
3, 10, 30, and 100 pc respectively.  Horizontal black lines indicate
the typical 5$\sigma$ sensitivity (left) and effective resolution
(right) of various telescopes used to observe the cloud.
Observational parameters from the following studies: CfA 1.2 m:
\citet{mwco}, Massachusetts-Stony Brook (MSB) survey: \citet{san86},
FCRAO Outer Galaxy Survey:
\citet{ogs}, NANTEN studies of the Magellanic Clouds: \citet{fu99},
BIMA studies of M33: \citet{ros03}, ALMA: web-based sensitivity calculator.}
\end{center}
\end{figure*}

To place these biases in the the context of real molecular cloud
studies, Figure \ref{DISTCOMP} shows the peak flux and angular size of
typical GMCs \citep[those of][]{srby87} as a function of distance. The
sensitivities and resolutions of a representative sample of molecular
cloud studies have been indicated as horizontal lines in these plots.
Distances to commonly observed objects have also been labeled.  Figure
\ref{DISTCOMP} demonstrates that most studies of extragalactic GMCs
are conducted where the clouds of interest are only marginally
resolved and are found at low sensitivity.  Even future observations of
GMCs in the Virgo cluster using the Atacama Large Millimeter Array
(ALMA) will be affected by resolution and sensitivity biases.

In this paper, we examine the effects of biases stemming from finite
spatial resolution, spectral resolution, and sensitivity in
molecular-line observations of GMCs. We recommend data analysis
methods to produce a standardized set of observed cloud properties
that account for these biases.  Most of the methods used in this paper
have been adopted piecemeal and {\it ad hoc} in previous studies.
Here, we endeavor to justify our choice of methodologies and to
synthesize various author's techniques for approaching the problems of
molecular cloud data analysis.  In Section \ref{MOMENTS}, we describe
a standardized method to measure three basic properties of an emission
distribution --- size, line width, and flux --- while accounting for
the sensitivity and resolution of a data set. In Section
\ref{PHYSICAL}, we discuss how these measurements can be transformed
into physical quantities --- radius, line width, luminosity, and
implied mass.  Finally, we consider the effects of using
interferometers to derive cloud properties in Section \ref{INTERF}.
The results in these three sections are applicable to all observations
of molecular clouds.  In contrast, the methods used for decomposing
emission into molecular clouds vary widely, and there is little basis
for favoring one method over another in all cases.  Hence, we defer a
presentation of our decomposition algorithm to the Appendix of the
paper, leaving only a brief discussion of the decomposition problem in
Section \ref{DECOMPSECT}.  We conclude the paper by exhibiting several
examples of the application of our standardized methods to previously
observed data and making recommendations for future observations
(Section \ref{APPLICATIONS}).

The methods described in this paper require a computer program to
apply. A documented software version of the decomposition and
measurement algorithms is available from the authors as an IDL
package.

\section{Measuring Molecular Cloud Properties \label{MOMENTS}}

This section describes how to derive the spatial size, line width, and
flux from a region of emission within a spectral line map (a ``data
cube'') while accounting for the finite sensitivity and resolution of
the data set.  We use moment methods \citep[e.g.~][]{srby87}, which
make full use of position and intensity information without assuming a
functional form for the cloud. They are therefore robust to
pathologies in the data {\it within} the cloud.

Moments are, however, sensitive to the inclusion of false emission
(noise) at the edge of a cloud. Including noise has the effect of
artificially increasing the values of the moment. Therefore the
methods outlined here should be employed in conjunction with careful
signal identification so that the calculations include as little noise
as possible. We assume throughout this section that the algorithm is
being applied to a distribution of real emission that we label a
``cloud'' (we discuss signal identification and decomposition in the
Appendix).

\subsection{Moment Measurements of Size, Line Width, and Flux}

This subsection describes how to apply moment methods to derive the
size, line width, and flux from a distribution of emission (a
``cloud'') within a position-position-velocity data cube. The data
cube consists of a number of pixels that have sizes of $\delta x$,
$\delta y$, and $\delta v$ in the two spatial dimensions and the
velocity dimension, respectively. The $i$th pixel in the data cube has
positions $x_i$ and $y_i$, velocity $v_i$, and brightness temperature
$T_i$. We assume that the cloud is contiguous and bordered by an
isosurface in brightness temperature of value $T_{edge}$, so that all
of the pixels in the cloud have $T > T_{edge}$ and the pixels outside
the cloud have $T < T_{edge}$ or are separated from the cloud by
emission with $T < T_{edge}$.

We begin by rotating the spatial axes so that the $x$ and $y$ axes
align with the major and minor axis of the cloud, respectively. We
determine the orientation of the major axis using principal component
analysis. We find the eigenvectors of the intensity-weighted
covariance matrix for the cloud,

\begin{equation}
\nonumber
\frac{1}{\sum_i T_i} 
\left[\begin{array}{ll}
\sum_i T_i \left( x_i - \bar{x} \right)^2 & 
\sum_i T_i \left( x_i - \bar{x} \right) \left( y_i - \bar{y}
\right) \\ 
\sum_i T_i \left( x_i - \bar{x} \right)
\left( y_i - \bar{y} \right) & 
\sum_i T_i \left( y_i - \bar{y} \right)^2 \\
\end{array}\right] \mbox{ .}
\end{equation}

\noindent In the equations above the sum $\sum_{i}$ runs over
all pixels within the cloud and $\bar{x}$ and $\bar{y}$ are the
intensity weighted mean positions within the cloud (defined below). We
define the new $x$ axis to lie along the eigenvector with the largest
eigenvalue --- the major axis of the cloud. The $y$ axis lies
perpendicular to the $x$ axis along the minor axis of the cloud. In
the discussion below, $x$ refers to position along the major axis and
$y$ refers to position along the minor axis. Rotating the axes in this
manner yields information about the axis ratio of the cloud and allows
a more careful deconvolution.  This method for determining the
position angle of molecular clouds has also been adopted by
\citet{koda}.

To measure of the size of the cloud, we compute the geometric mean of
the second spatial moments along the major and minor axis. This is
$\sigma_{r}$, the root-mean-squared (RMS) spatial size:

\begin{equation}
\sigma_{r} (T_{edge}) = \sqrt{\sigma_{maj} (T_{edge})~\sigma_{min} (T_{edge})}
\end{equation}

\noindent where $\sigma_{maj} (T_{edge})$ and $\sigma_{min}
(T_{edge})$ are the RMS sizes (second moments) of the intensity
distribution along the two spatial dimensions.  We adopt this
particular functional form since it has been used in previous
observational studies \citep{srby87} and explored in depth by
\citet{bert92} with respect to inclination, aspect ratio, and
virialization. We calculate $\sigma_{maj} (T_{edge})$ and
$\sigma_{min} (T_{edge})$ by:

\begin{eqnarray}
\sigma_{maj} (T_{edge}) &= &\sqrt{\frac{\sum_{i}^{cloud} T_i \left[ x_i - \bar{x}
  (T_{edge}) \right]^2}{\sum_{i}^{cloud} T_i}}, \mbox{ where} \\
\bar{x}
  (T_{edge}) &= &\frac{\sum_{i}^{cloud} T_i x_i}{\sum_{i}^{cloud} T_i}
  \mbox{~and} \\ 
\sigma_{min} (T_{edge}) & = & \sqrt{\frac{\sum_{i}^{cloud} T_i
  \left[ y_i - \bar{y} (T_{edge}) \right]^2}{\sum_{i}^{cloud} T_i}} \mbox{, where} \\
 \bar{y} (T_{edge}) & = & \frac{\sum_{i}^{cloud} T_i
  y_i}{\sum_{i}^{cloud} T_i} \mbox{.}
\end{eqnarray}

\noindent In the equations above the sum $\sum_{i}^{cloud}$ runs over
all pixels within the cloud. We have written each of the moments as a
function of $T_{edge}$ because changing the isosurface that defines
the boundary of the cloud ($T_{edge}$) will change the set of pixels
included in the sum and therefore the values of the moments. Note that
$\sigma_{r}$ is not the RMS distance ($d = \sqrt{x^2 + y^2}$) from the
center of the cloud. Rather it is the analogous to the RMS size of the
cloud along an arbitrarily chosen axis. Thus, $\sigma_{r} = \sigma_{x}
= \sigma_{y}$ for a perfectly round cloud, while the RMS distance from
the center for such a distribution is larger, $\sigma_{d} = \sqrt{2}
\sigma_{x} > \sigma_{x}$. Also note that $\sigma_{maj} / \sigma_{min}$
is the axis ratio of the cloud and will be $\sim 1$ for round clouds
and $\gg1$ for elongated or filamentary clouds.

We calculate the velocity dispersion, $\sigma_v(T_{edge})$ in the same
manner as the size:

\begin{eqnarray}
\sigma_v (T_{edge})& = &\sqrt{\frac{\sum_{i}^{cloud} T_i \left[ v_i -
\bar{v} (T_{edge}) \right]^2}{\sum_{i}^{cloud} T_i}} \mbox{, where} \\
\bar{v} (T_{edge})& = & \frac{\sum_{i}^{cloud} T_i v_i}{\sum_{i}^{cloud}
T_i} \mbox{ .}
\end{eqnarray}

\noindent The sums again run over all emission in the cloud. For a
Gaussian line profile, such as that found for most clouds, the
full-width half-maximum (FWHM) line width, $\Delta V (T_{edge})$ will
be related to $\sigma_v (T_{edge})$ by

\begin{equation}
\Delta V (T_{edge}) =\sqrt{8\ln(2)}~\sigma_v (T_{edge}) \mbox{.}
\end{equation}

\noindent Finally, we calculate the flux of the cloud,
$F_{\mathrm{CO}} (T_{edge})$ using the zeroth moment:

\begin{equation}
F_{\mathrm{CO}} (T_{edge}) = \sum_i T_i~\delta v~\delta x~\delta y
\mbox{.}
\end{equation}

\noindent If $\delta x$ and $\delta y$ are in units of arcseconds,
$\delta v$ in km s$^{-1}$, and $T_i$ in K, then the resulting flux
will have units of K km s$^{-1}$ arcsecond$^{2}$.

\subsection{Correcting for the Sensitivity Bias}
\label{EXTRAPSECT}

The sensitivity of a dataset influences the cloud properties derived
from that data, a fact that we have emphasized in the previous section
by explicitly writing the moments as functions of $T_{edge}$, the
cloud boundary (usually set by the signal-to-noise ratio of the
data). Figure \ref{EXTRAPMOMS} shows the variation of spatial size,
line width, and flux as a function of sensitivity for a bright cloud
in M33. The data for this cloud shows a substantial {\em sensitivity
bias}; all of the derived properties are strong functions of the
boundary isosurface ($T_{edge}$). In order to compare data sets with
different sensitivities, one must account for this bias. In this
section we describe a method to do this by extrapolating the measured
properties of a cloud---$\sigma_{maj} (T_{edge})$, $\sigma_{min}
(T_{edge})$, $\sigma_v (T_{edge})$, and $F_{\mathrm{CO}}
(T_{edge})$---to those we would expect to measure for a cloud within a
boundary isosurface of $T_{edge} = 0$ Kelvin (i.e., perfect
sensitivity).

We estimate the values of the moments at $T_{edge}=0$~K by
extrapolating from higher values of $T_{edge}$.  This technique was
originally suggested for inferring total cloud areas by \citet{bt80}
and applied to molecular cloud properties by \citet{syscw}.  We
calculate each of the moments for a sample of boundary temperatures,
$T_{edge}$, ranging from near the peak temperature of the cloud to the
lowest boundaries allowed by the data. Thus, we measure the variations
of the moments as a function of the boundary temperature, $T_{edge}$,
within the cloud (this is how we constructed the plot shown in Figure
\ref{EXTRAPMOMS}). Below, we assume that we have measured each of the
four moments for values of $T_{edge}$ ranging from $T_{min}$ (the
minimum allowed by the data, that is the sensitivity limit) to
$T_{max}$ (near the peak temperature of the cloud).

\begin{figure}
\begin{center}
\plotone{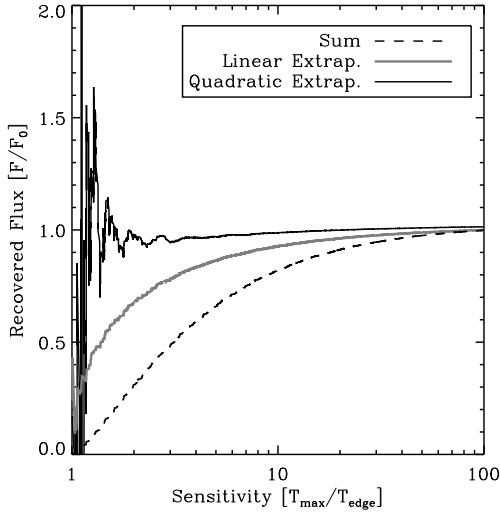} \figcaption{\label{FLUXFIG} The recovered
flux as a function of sensitivity, $T_{peak}/T_{edge}$. The three
lines show the fraction of the total flux measured for a Gaussian
cloud using no sensitivity correction (dashed line), a linear
extrapolation to perfect sensitivity (gray line), and a quadratic
extrapolation (black line). The quadratic extrapolation does an
excellent job of recovering the correct value of the flux down to low
sensitivities. The uncorrected sum and the flux corrected by a linear
extrapolation both show a substantial sensitivity bias.}
\end{center}
\end{figure}

We estimate the value of the moments at $T_{edge} = 0$~K by performing
a weighted, linear least-squares fit to the measured moments.  As an
example, we consider $\sigma_{maj} (T_{edge})$.  The data are modeled as
\begin{equation}
\sigma_{maj} (T_{edge}) = m~T_{edge} + \sigma_{maj} (0~\mbox{K})
\end{equation}
and the fit determines the extrapolated moment, $\sigma_{maj}
(0~\mbox{K})$.  For the fit, each pair of data $\{T_{edge},
\sigma_{maj}(T_{edge})\}$ is assigned a weight proportional to the
number of data in the cloud with $T>T_{edge}$, so that measurements of
the moment using more data are weighted more heavily. Practically,
this means that points to the left in Figure \ref{EXTRAPMOMS} have
higher weights than those to the right. We use this linear
extrapolation for $\sigma_{maj}$, $\sigma_{min}$, and $\sigma_v$, but
we find that a quadratic extrapolation (including a $T_{edge}^2$ term)
gives better results for the zeroth moment, $F_{\mathrm{CO}}$ (though
we revert to a linear extrapolation when the extrapolated flux is
lower than the measured flux).  We plot the flux of a Gaussian cloud
as a function of $T_{edge}$ in Figure \ref{FLUXFIG} for an uncorrected
zeroth moment and the linear and quadratic extrapolations to
illustrate the appropriateness of the quadratic fit to the zeroth
moment. At very low sensitivities (signal-to-noise ratios near unity),
the quadratic extrapolation is very noisy, but for signal-to-noise
ratios of $2$ or better it does a dramatically better job of recovering
the true flux of the cloud ($F/F_0 = 1$) than either the linear
extrapolation or no extrapolation.

Figure \ref{EXTRAPMOMS} also shows these extrapolations for a bright
cloud in M~33.  The result of this extrapolation is a set of four
moments --- $\sigma_{maj} (0~\mathrm{K})$, $\sigma_{min} (0
~\mathrm{K})$, $\sigma_v (0~\mathrm{K})$, and $F_{\mathrm{CO}} (0
~\mathrm{K})$ --- that correspond to those we would expect to measure
given infinite sensitivity. The values of these moments should be
directly comparable even among datasets with different sensitivities
(values of $T_{min}$).

Note that diffuse emission surrounding a GMC may confuse this
method. If one data set is measured with sensitivity sufficient to
detect diffuse emission surrounding a GMC, while another lacks the
sensitivity to do so then this approach may not be sufficient to
correct for the sensitivity bias. This problem may be particularly
acute when comparing Galactic GMCs observed with very good sensitivity
to extragalactic clouds with worse signal-to-noise
ratios. Interferometric data ``resolves out'' emission significantly
more extended than the synthesized beam (see \S\ref{INTERF}),
representing another bias against detecting diffuse
emission. \citet[][]{polk},
\citet[][]{blitz85}, \citet[][]{ros03}, and \citet[][]{ler05} find
evidence for diffuse emission surrounding GMCs in the Milky Way and
the Local Group Galaxies M~31, M~33, and IC~10, respectively.

\subsection{Correcting for the Resolution Bias}
\label{BEAMDCSECT}

Any astronomical data set represents the convolution of the intensity
of the source with the profile of the instrument used to observe
it. Care must therefore be taken in measuring sizes and line widths
when the extent of the intensity distribution is comparable to the
instrumental profile. In a typical spectral line data cube two
profiles are important: the spatial beam and the width of a velocity
channel. In this section, we describe simple corrections to account
for the effects of finite spatial and spectral resolution.

We ``deconvolve'' the spatial beam from the measured cloud size by
subtracting the RMS beam size, $\sigma_{beam}$, from the extrapolated
spatial moments, $\sigma_{maj} (T_{edge} = 0~\mathrm{K})$ and
$\sigma_{min} (T_{edge} = 0~\mathrm{K})$, in quadrature --- an approach
that is exact for Gaussians. The deconvolved second moment is given
by:

\begin{equation}
\sigma_{r,dc} = \sqrt{ [ \sigma_{maj}^2 \left(0~\mathrm{ K} \right)
- \sigma_{beam}^2 ]^{1/2} ~ [ \sigma_{min}^2 \left(0~\mathrm{
K} \right) - \sigma_{beam}^2 ]^{1/2}}~\mbox{ ,}
\end{equation}

\noindent where $\sigma_{maj} \left(0~\mathrm{ K} \right)$ and
$\sigma_{min} \left(0~\mathrm{K} \right)$ are extrapolated to the 0
Kelvin isosurface as described in \S\ref{EXTRAPSECT}.  This
extrapolation is necessary to make this deconvolution valid:
subtracting the full $\sigma_{beam}$ from the spatial moment measured
for only part of the cloud will lead to an overcorrection and thus to
an underestimate of the cloud size. Measuring the spatial size along
the minor axis is also necessary to ensure that the cloud is indeed
resolved in all dimensions. This is an advantage of the choice of axes
(major/minor) described above. With sufficient signal-to-noise, this
method of deconvolution provides a robust measurement of cloud size
even for marginally resolved clouds.

Instrumental resolution also affects the measured line
width. Spectrometers measure the average intensity across a channel,
rather than sampling the intensity at the center (nominal frequency)
of that channel. When the width of the spectral line under
consideration is comparable to the bandwidth of a single channel, the
line strength varies significantly across an individual channel. In
this case, the average value may differ substantially from the value at
line center. The output of the spectrometer is thus a convolution of
the true spectral profile with the profile of an individual
channel. We account for this potential bias towards higher line widths
by a simple deconvolution of the channel width from the extrapolated second
moment:

\begin{equation}
\sigma_{v,dc}=\sqrt{{\sigma_{v}^2 \left(0~\mathrm{
      K}\right)-\frac{\Delta V^2_{chan}}{2~\pi}}}
\end{equation}
where $\sigma_{v} \left(0~\mathrm{ K} \right)$ is the second moment of
the cloud in the $v$ dimension extrapolated to 0 Kelvin as described
in \S \ref{EXTRAPSECT} and $\Delta V_{chan}$ is the width of a
velocity resolution element. Although the channel profiles are usually
square in shape and not Gaussian, we simplify the deconvolution by
approximating the channel shape as a Gaussian with integrated area
equal to that of a square channel with width $\Delta V_{chan}$.  For
such a Gaussian, $\sigma_{chan}=\Delta V_{chan}/\sqrt{2~\pi}$.

\subsection{Comparison with Other Methods}
\label{compare}
We use extrapolated moments to measure GMC properties rather than
employing an established method from the literature. In this section,
we justify our choice by comparing several methods of measuring GMC
properties. We focus on the performance of these methods at marginal
resolution and low signal-to-noise, conditions typical of
extragalactic GMC observations.

Determining the radius of a cloud is particularly difficult because
GMCs are often asymmetrical with poorly defined boundaries. Several
authors have devised methods to return a single characteristic size
for complicated emission distributions. The intensity-weighted second
moments in the spatial directions have been used in many studies
\citep[e.g.~][]{srby87}, but are sensitive to both noise and
convolution effects. The other commonly used method
\citep[e.g.~][]{clumpfind,hc01} is to infer the radius based on the
area of the cloud:
\[R_e = \sqrt{\frac{A-A_{pt}}{\pi}}.\]
Here $A_{pt}$ is the area of a point source that has been convolved
with a beam and measured with the same signal-to-noise as the emission
in the map.  Finally, \citet{ws90} and \citet{tay99} adopted the size
of the cloud as the mean of the deconvolved FWHMs of the emission
distribution along two perpendicular directions. 

\begin{figure*}
\begin{center}
\plotone{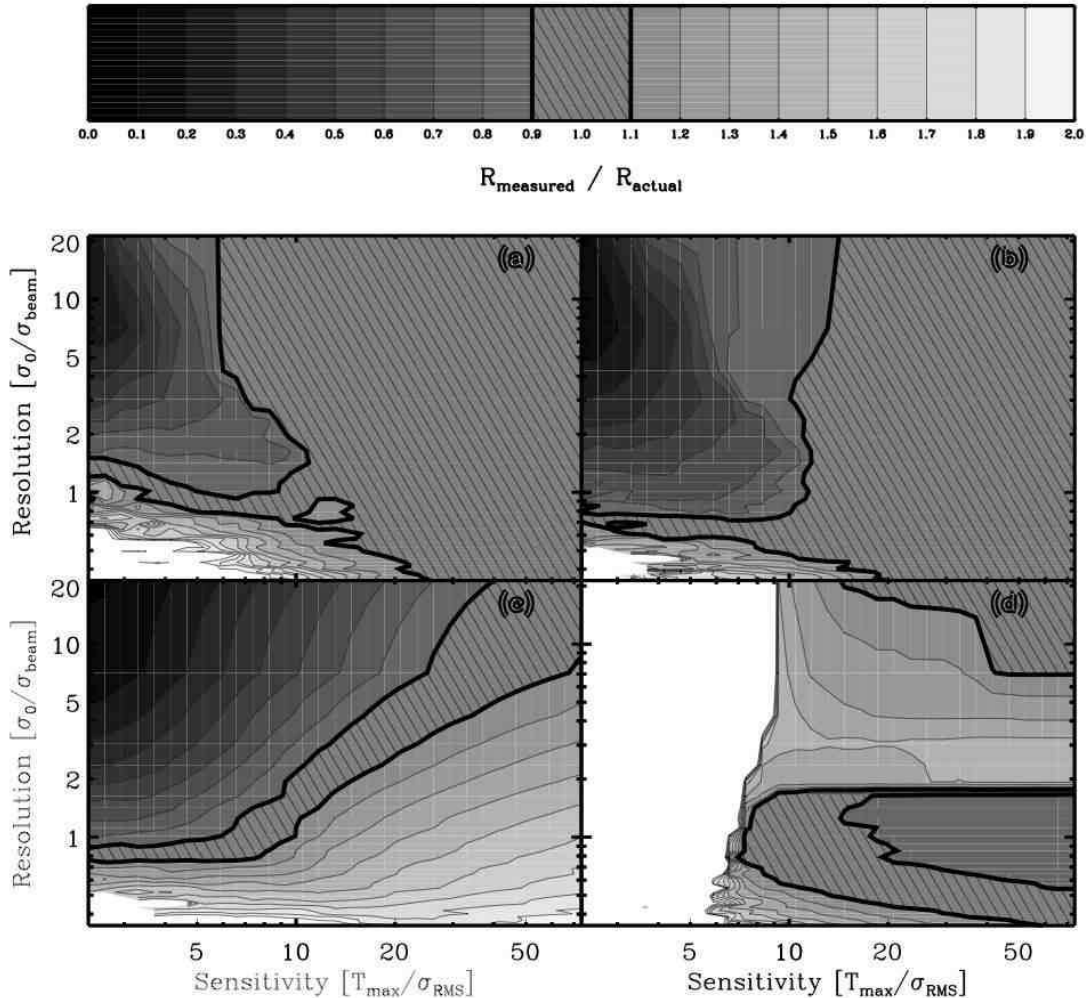} \figcaption{\label{radplot} The recovered radius
as a function of sensitivity ($T_{max}/\sigma_{RMS}$) and resolution
($\sigma_0/\sigma_{beam}$). The four panels show the radius recovered
by (a) the extrapolated moment method (this paper), (b) the
uncorrected moment method, (c) the area method, and (d) the FWHM
method. The hashed region indicates the where each algorithm obtains a
result within 10\% of the original value. The extrapolated moment
accurately recovers the radius over a large zone of parameter
space. Of particular importance for extragalactic measurements, it
performs better at low signal-to-noise than the other three methods.}
\end{center}
\end{figure*}

We compare these three methods to the extrapolated moment method
presented above across a range of resolutions and sensitivities. We
measure the size of a Galactic GMC using each method after convolving
it to a desired resolution and adding noise to produce a particular
signal-to-noise ratio. For the data, we use the $^{12}$CO data from
the Rosette molecular cloud \citep{bs86}, which we clip at
2$\sigma_{RMS}$ (the RMS noise in the original data set) and integrate
in the velocity dimension to produce a map of integrated
intensity. For a range of sensitivities and resolutions, we convolve
this map with a Gaussian beam and add noise.  We measure the size of
the cloud in 100 realizations of the noise for each such \{resolution,
sensitivity\} pair using (a) the extrapolated moment method, (b) the
moment of the data without extrapolation, (c) the area method and (d)
the FWHM method.  Wherever possible we corrected for the effects of
beam convolution and signal-to-noise for each of the methods.  The
results of the analysis are shown in Figure \ref{radplot}.

Figure \ref{radplot} shows the recovered radius as a function of
resolution and sensitivity, with the ``true'' radius defined as that
measured at very high sensitivity and very good resolution (i.e. in
the original data, the top right corner of each plot). The hashed
region of parameter space shows the range of parameters over which
each algorithm recovers a radius within 10\% of the true value. The
extrapolated moment method has the largest hashed region and so is
remarkably robust, recovering the true radius over a large range of
sensitivities and resolutions. Only at low sensitivity
($T_{max}/\sigma_{RMS}<5$) and marginal resolution
($\sigma_{beam}\gtrsim \sigma_0$), does the derived radii depart
systematically from the true radius. Notably, the extrapolated second
moment performs quite well at signal-to-noise ratios from 5 to 10 (in
the integrated intensity map), values typical of extragalactic CO data
sets. By contrast, the uncorrected moment method (panel b in Figure
\ref{radplot}) underestimates the size at low signal-to-noise since
(by construction) the uncorrected moment does not account for emission
below the noise level. Similarly, the area method (panel c) shows
systematic variation at both low signal-to-noise (where emission drops
below the noise level) and low resolution (where the convolved area of
the emission distribution grows disproportionately because of the
filamentary nature of the cloud).  Finally, the FWHM method (panel d)
shows large systematic variations since it depends only on the
location of the FWHM contour and not on the remainder of the emission
distribution.

The region of systematic underestimation at low signal-to-noise but
reasonable resolution shows the effects of missing the diffuse
emission mentioned in \S\ref{EXTRAPSECT}. The Rosette includes more CO
emission at low intensities than the extrapolated moment predicts from
the high intensity data. As a result, when that diffuse emission is
not included in the measurement, the algorithm underestimates the true
radius of the GMC. This effect is seen panels (a) and (b) --- the
extrapolated and uncorrected second moments --- and is more pronounced
in the uncorrected second moment, panel (b).

\begin{figure*}
\begin{center}
\plotone{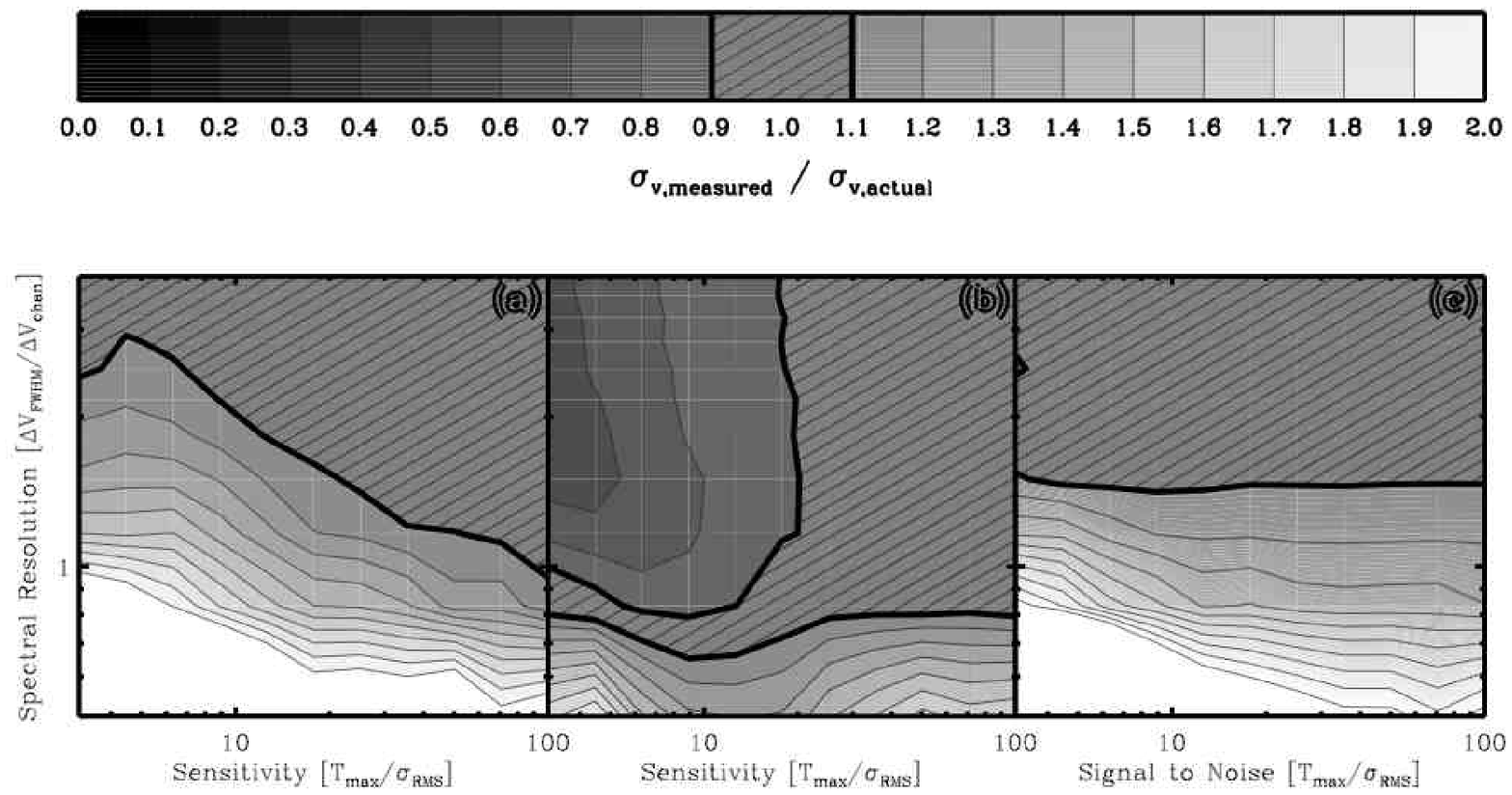} \figcaption{\label{dvplot} The measured line width
as a function of sensitivity ($T_{max}/\sigma_{RMS}$) and spectral
resolution ($\Delta V_{FWHM}/\Delta V_{chan}$). The three panels show
the line width of a Gaussian with known properties measured using (a)
the extrapolated moment method (this paper), (b) an uncorrected second
moment and (c) a Gaussian fit to the resulting spectrum. The hashed
region indicates the zone of parameter space where each algorithm
measures a line width within 10\% of the actual value. The
extrapolated moment provides a good measurement of the line width,
remaining comparable to a direct fit even in the limit of low
signal-to-noise and spectral resolution. Because the Gaussian fit is
provides poor results for non-Gaussian line profiles, we prefer the
extrapolated moment method.}
\end{center}
\end{figure*}

We perform a similar experiment on recovering the line width of an
emission line in noisy data.  We measure the recovered line width of a
Gaussian line of known width using three methods (a) the extrapolated
moment method (b) an uncorrected second moment and (c) a Gaussian fit
to the line.  For a range of signal-to-noise levels
($T_{max}/\sigma_{RMS}$) and channel widths $\Delta V_{chan}$ we
measure the recovered line width relative to the known line width.
The mean values of the recovered line for 1000 realizations of the
noise are plotted in Figure \ref{dvplot}.  The extrapolated moment
does not show the systematic variation with signal-to-noise seen in
the uncorrected moment.  The extrapolated moment is nearly as robust a
measure as the Gaussian fit for a perfectly Gaussian line and will
prove superior if the line is not Gaussian.  Robust recovery of the
line width using any method requires $\Delta V_{line} / \Delta
V_{chan} > 2$.

\subsection{Assessing Errors in GMC Properties}

The formal uncertainty associated with each moment measurement is
quite small. Cloud identification and extrapolation represent larger
sources of uncertainty, but their effects are difficult to assess
formally.  We use bootstrapping methods to estimate the uncertainties
in our measurements of cloud properties.  The bootstrapping method
determines errors by generating several trial clouds from the original
cloud data.  A trial cloud is generated by considering the cloud to be
a collection of data $\{x_i,y_i,v_i,T_i\}$ for $i=1\dots N$, the
number of points in the cloud.  The data are sampled for $N$ random
values of $i$, allowing for $i$ to be repeated.  The properties of the
cloud are measured for each trial cloud.  We estimate the uncertainty
from the variance of the cloud properties derived from these resampled
and remeasured data sets. The final uncertainty in each property is
the standard deviation of the bootstrapped values scaled up by the
square root of the oversampling rate. The oversampling rate, which is
usually equal to the number of pixels per beam, accounts for the fact
that not all of the data in each cloud are independent. For many
interferometric data sets this is an important effect, since these
data can have $10$ or more pixels per beam.

We compare the uncertainties produced by the bootstrapping to those
derived from repeatedly adding noise to and then reanalyzing a data
set. We use the bright cloud in M~33 shown in Figure
\ref{EXTRAPMOMS}. We conduct 100 realizations of the data plus random
noise. The resulting uncertainties in $\sigma_{maj}$, $\sigma_{min}$,
$\sigma_v$, and the flux are $3\%$, $2\%$, $3\%$, and
$3\%$. Repeatedly bootstrapping the same data set (adjusted to have
the same final noise level) yields average uncertainties of $9\%$,
$9\%$, $11\%$, and $5\%$. The bootstrapping estimates are higher for
this bright cloud because they reflect both the formal uncertainty and
the robustness of the result to the removal of a given piece of
data. In the low signal-to-noise regime, the values for the two
methods converge as noise dominates the uncertainty derived from
bootstrapping --- for example, performing the same experiment in a
dimmer M~33 cloud with $1/4$ the luminosity of the bright cloud and
comparable noise, bootstrapping yields errors of $31\%$, $33\%$,
$32\%$, and $35\%$ in the four moments while repeated realizations
produces scatters of $15\%$, $27\%$, $30\%$, and $40\%$.

The bootstrapping method produces a robust, believable estimate of the
uncertainty in the measurement of the properties of a particular,
defined cloud. It does not account for uncertainties in the assignment
of emission to a cloud either as a result of noise or choice of
algorithm. These uncertainties are more systematic than random in
nature and may be best assessed by analyzing the emission distribution
using several methods. The bootstrapping estimate may be treated as an
accurate estimate of the uncertainties in the results {\em given} that
one adopts the methods presented in this paper.

\section{Deriving Physical Quantities from Moment Measurements
  \label{PHYSICAL}}

In this section, we outline how to use the measured size, line width,
and flux to calculate several physical quantities of interest: the
effective spherical radius, the virial mass, and the luminous
mass. Throughout this section we assume that clouds can be described
as self-gravitating spheres with density profiles $\rho \propto
r^{-1}$ and negligible support from magnetic fields or confinement by
external pressure.

We assume below that the data consists of observations of the \co\
\jone\ transition, in units of brightness temperature, but the method
is readily adaptable to analogous data sets.

\subsection{The Spherical Radius}

We define a factor $\eta$ that relates the one-dimensional RMS size,
$\sigma_r$, to the radius of a spherical cloud $R$: $R=\eta \sigma_r$.
It is possible to derive an estimate for $\eta$ based on spherical
cloud of radius $R$ with a density profile $\rho \propto r^{-\beta}$.
In this model,
\begin{eqnarray}
\sigma_r^2 &=& \frac{\int_0^R dr \int_0^{2\pi} d\theta \int_0^{\pi} d\phi~x^2
\rho_0 r^{-\beta} r^2 \sin \phi}{\int_0^R dr
\int_0^{2\pi} d\theta \int_0^{\pi} d\phi~ \rho_0 r^{-\beta} r^2 \sin \phi} \\
\sigma_r^2 &=& \frac{1}{3}~\frac{3 - \beta}{5 - \beta} R^2 \\
\mathrm{so~that~} \eta &=& \sqrt{3~\frac{5 - \beta}{3 -
\beta}}
\end{eqnarray}

\noindent For a cloud with $\beta = 1$, $\eta = \sqrt{6}$, somewhat
higher than the empirical correction of $3.4/\sqrt{\pi}$ derived by
Solomon et al. (1987).  The difference arises, in part, from using
$^{12}$CO as a density tracer.  Since $^{12}$CO emission saturates in
dense regions and vanishes from low density regions, the apparent
density profile in $^{12}$CO is shallower than the true density
profile.  Hence, an appropriate value of $\eta$ likely falls in the
range between $3.4/\sqrt{\pi}$ and $\sqrt{6}$.  For tracers like
$^{13}$CO with higher critical densities, a different value of $\eta$
may be appropriate.  Since comparison to this ``anchoring'' data set
may be more important than adopting a self-consistent --- but grossly
oversimplified --- model for a cloud, we recommend the Solomon et
al. (1987) definition of the cloud radius, $R \approx 1.91
\sigma_{r}$.  Note, that adjusting the definition of the radius
renders the virial mass formula we present below inexact.

\subsection{The Spatial Size, Line Width, Luminosity, and Mass}

We convert the cloud properties $\sigma_r ( 0~\mathrm{K})$, $\sigma_v
( 0~\mathrm{K})$, and $F_{\mathrm{CO}} ( 0~\mathrm{K})$ to the physical
quantities $R$, $\Delta V$, and $L_{\mathrm{CO}}$. For a cloud at a distance
of $d$ (in parsecs), the physical radius will be

\begin{equation}
R[\mathrm{pc}] = \frac{R ( 0~\mathrm{K}) [\mathrm{arcsec}]}{3600}
\times \frac{\pi}{180} \times d [\mathrm{pc}]~\mathrm{,}
\end{equation}

\noindent the FWHM line width will be

\begin{equation}
\Delta V = \sqrt{8\ln (2)}~\sigma_v ( 0~\mathrm{K})~\mathrm{,}
\end{equation}

\noindent and the luminosity of the cloud, $L_{\mathrm{CO}}$, will be

\begin{eqnarray}
\nonumber
L_{\mathrm{CO}} [\mathrm{K~km~s}^{-1}\mathrm{~pc}^2] &=&
    F_{\mathrm{CO}}(0~\mathrm{K}) [\mathrm{K~km~
    s}^{-1}~\mathrm{arcsec}^2] \\
\nonumber
&\times& (d[\mathrm{pc}])^2\\
&\times& \left(\frac{\pi}{180\cdot 3600}\right)^2.
\end{eqnarray}

\noindent A particular CO luminosity, $L_{\mathrm{CO}}$, implies a mass of
molecular gas, $M_{\mathrm{Lum}}$, of

\begin{eqnarray}
\nonumber
{M_{\mathrm{Lum}}}~[{M}_{\odot}]& = &\frac{X_{\mathrm{CO}}}{2
  \times 10^{20} [\mathrm{cm}^{-2}/(\mathrm{K~km~s}^{-1})]} \times
  4.4~L_{\mathrm{CO}}\\
&\equiv& 4.4~X_2~L_{\mathrm{CO}}
\end{eqnarray}

\noindent where $X_{\mathrm{CO}}$ is the assumed CO-to-H$_2$
conversion factor. This calculation includes a factor of 1.36 (by
mass) to account for the presence of helium. Including helium is
important to facilitate comparison with the virial mass, which should
reflect all of the gravitating mass in the cloud.  We have adopted a
fiducial value of the CO-to-H$_2$ conversion factor of
$X_{\mathrm{CO}}=2\times 10^{20}\mbox{ cm}^{-2} (\mbox{K km
s}^{-1})^{-1}$ and express changes relative to this value in terms of
the parameter $X_2$.

\subsection{The Virial Mass}

We compute the virial masses under the assumption that each cloud is
spherical and virialized with a density profile described by a
truncated power law of the form $\rho \propto r^{-\beta}$ with no
magnetic support or pressure confinement. As with the spherical radius
correction, the exact density profile of the cloud will affect the
correct form of the virial theorem mass. For $\beta = 1$, the virial
mass is given by the formula (Solomon et al. 1987):

\begin{equation}
{M_{\mathrm{VT}}} = 189~M_{\odot}~\Delta V^2 \, R
\end{equation}

\noindent and more generally by

\begin{equation}
{M}_{\mathrm{VT}} = 125~M_{\odot}~\frac{5-2\beta}{3-\beta}~\Delta V^2 \,
R~\mathrm{,}
\end{equation}

\noindent where $\Delta V$ is the FWHM velocity line width in \kmpers,
$R$ is the radius in pc, and the cloud has a density profile of $\rho
\propto r^{-\beta}$.

Clouds exhibit a range of non-spherical geometries and may be
supported by magnetic fields or confined by pressure. Therefore, the
studying the virial parameter may be more useful than the virial mass
itself. The virial parameter is a constant of order unity that
characterizes deviations from the virial theorem applied to a
non-magnetic cloud with no external pressure and constant density.
Following \citet{mckee-vt}, we define the virial parameter as
\begin{equation}
\alpha = \frac{5 \sigma_v^2 R}{M_{\mathrm{Lum}} G} =  
\frac{5 \eta \sigma_v^2 \sigma_r}{(4.4 X_2 L_{\mathrm{CO}})
G}~\mbox{.}
\end{equation}
Larger-than-unity virial parameters can result from pressure
confinement, while $\alpha<1$ may result from significant magnetic
support. Incorrect values of the CO-to-H$_2$ conversion factor may
skew the result in either direction.  Finding $\alpha<2$ means that the
clouds are gravitationally bound in the absence of significant
magnetic support.

\section{Measuring Cloud Properties from Interferometer Observations}
\label{INTERF}

Millimeter-wave interferometers are required to resolve even the most
massive molecular clouds in galaxies beyond the Magellanic clouds (see
Figure \ref{DISTCOMP}).  Unfortunately, interferometers are not
sensitive to spatial frequencies outside the limited region of the
$(u,v)$ plane that they sample. Practically, this means
interferometers do not measure the total flux from the emission
distribution; and structures are resolved out, usually on large angular
scales that correspond to small separations in the $(u,v)$ plane.
Ideally, interferometer observations are combined with single-dish
observations that supply the missing information.  In practice such
observations are conducted infrequently and the unknown total flux and
short-spacing information is estimated using deconvolution algorithms
such as CLEAN or maximum entropy
\citep[see][and references therein]{tms}. 

\citet{she00} simulate the results of using only an interferometer to
observe Galactic GMCs as if these well-studied clouds were located in
M31.  They find that interferometers experience significant (50\%)
flux loss for their simulated observation, primarily from extended
emission around clouds. However, they find that the flux loss does not
change the size and line width of the cloud. \citet{hel02} examine the
recovery of large-scale flux distributions from interferometer
measurements in more depth and explore the effectiveness of
deconvolution algorithms at low signal-to-noise. They find that
deconvolution algorithms recover flux nonlinearly at low
sensitivities, finding much less flux at low sensitivities than one
would expect. Since much of the data on extragalactic GMCs have low
signal-to-noise, this may represent an important bias.

We assess the effects of interferometric biases on the methods
presented here by extending the method of \citet{she00}. We use
$^{12}\mathrm{CO}$ observations of three galactic GMCs: the Orion
molecular complex \citep{wil05}, the Rosette Molecular Cloud
\citep{bs86}, and an excerpt from the Outer Galaxy Survey of
\citet{hc01} which contains the molecular clouds associated with the
W3/W4/W5 \ion{H}{2} regions\footnote{A map of the original Orion data
and typical maps at low resolution and sensitivity appear in the
Appendix.}. We simulate observing these three molecular complexes in
M31 ($D=770$ kpc) with the BIMA interferometer.

We Fourier transform each plane of each data set into the $(u,v)$
domain and resample the data along the $(u,v)$ tracks that would be
sampled by BIMA observations of the data provided the GMCs were in
M31. The $(u,v)$ coverage reflects typical observing strategies for
extragalactic clouds, such as interleaving observations of the source
with calibrators and other sources. We add thermal noise and phase
noise to the $(u,v)$ data, including a phase noise component with a
magnitude that depends on the length of the baseline.  We adjust the
scale of the thermal and phase noise to produce the desired peak
signal-to-noise (we find our results depend only weakly on whether the
noise is thermal or phase). We invert the resulting $(u,v)$ data using
the MIRIAD software package \citep{miriad} producing maps separated by
2 km s$^{-1}$, and then we deconvolve the dirty maps using a CLEAN
algorithm that terminates at the 2$\sigma_{RMS}$ level. For each trial
cloud, we then calculate the cloud properties using the methods of
\S\S \ref{MOMENTS} and \ref{PHYSICAL}.

For comparison, we compute cloud properties using the same procedure
to simulate single-dish observations with signal-to-noise and
effective resolution identical to the mock interferometer data.  We
generate the mock single-dish observations by sampling the transformed
image data for an equal number of $(u,v)$ points as the interferometer
data, but the points are normally distributed in the $(u,v)$ plane and
one point is forced to lie at $(0,0)$ thus sampling the total
power. The width of the $(u,v)$ point distribution is chosen to give a
beam size similar to that of the mock interferometer data.  Random
thermal and phase noise is added to these data in the same fashion as
for the interferometer data.  Then the data are inverted using natural
weighting and deconvolved in the exact same fashion as the
interferometer data (though the deconvolution step has little effect).
Again, we extract cloud properties using the methods of
\S\S\ref{MOMENTS} and \ref{PHYSICAL}.  


With these simulations, we compare the cloud properties derived from
interferometer data to single-dish data that are equivalent in every
other fashion, thereby isolating the effects of interferometers on the
derived properties.  The additional biases imposed by limited
resolution and sensitivity are discussed separately in
\S\ref{compare}.  Here we focus on mock BIMA observations of the three
molecular complexes using three antenna configurations: the C array
(extended), the D (compact) array, and a combination of C and D array
\citep[see][for details on the configurations]{wright-configs}. 
The synthesized beam sizes for these configurations are $14.4''$,
$6.1''$ and $8.8''$ for the D, C and C+D hybrid array observations
respectively, corresponding in turn to 54, 22 and 33 parsecs at
770~kpc. We conduct 10 sample observations for each cloud in each
array at a range of sensitivities ranging from
$T_{peak}/\sigma_{RMS}=3$ to $>100$.

\begin{figure*}
\begin{center}
\plotone{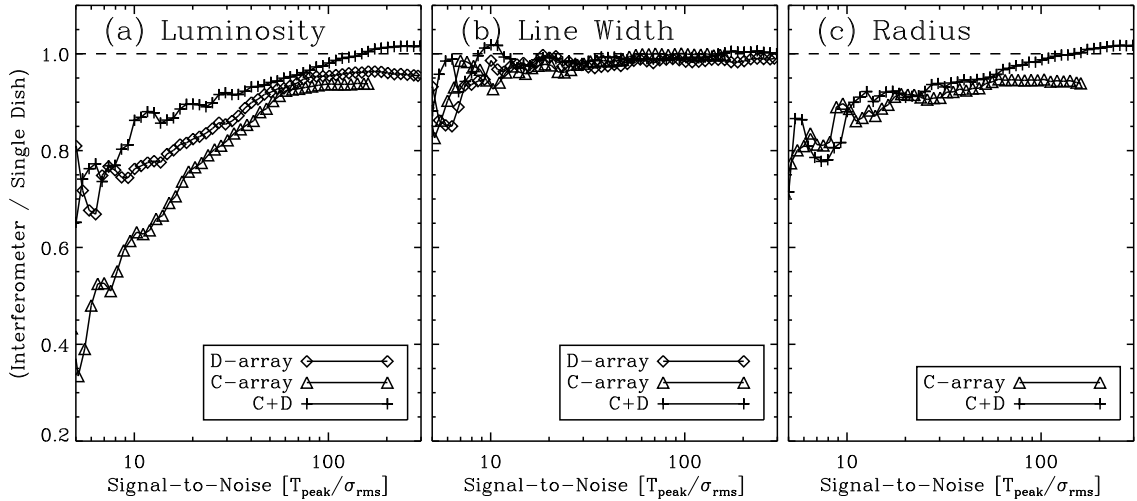}
\caption{\label{interf_props} The effects of interferometer observations 
on the measurement of cloud properties.  Each panel shows the
variation of cloud properties as a function of signal-to-noise for
interferometer observations.  The value of each property (luminosity,
line width, radius) is normalized by the value that is measured for
single dish observations at the same resolution and sensitivity as
that of the interferometer.  The Figure shows the behavior of cloud
properties for simulated BIMA observations of the Orion molecular
complex as if the clouds were located in M31.  No D-array curve is
plotted for radius measurements since Orion would be unresolved in
M31.  The recovery of cloud properties is a function of both
sensitivity and the $(u,v)$ coverage of the interferometer.}
\end{center}
\end{figure*}

Figure \ref{interf_props} shows the properties recovered by mock
observations of the Orion molecular complex for each array and a range
of sensitivities. The values of each properties are normalized by the
value recovered by mock single-dish observations at the same
sensitivity and resolution.  Thus, the only difference between the
four sets of properties (C, D, C+D, and single dish) is the $(u,v)$
coverage of the simulated observations.  We find that the derived
properties from mock single-dish observations follow the same behavior
as the simulations in \S\ref{compare}.  Thus, it is possible to
decouple the two sets of biases -- those arising from marginal
resolution and sensitivity and those arising from using
interferometers -- and examine only the latter.  We plot the results
for Orion because these observations show the most dramatic variation
of the three complexes studied, but the results are qualitatively the
same for all three data sets. Orion is the most sensitive of the three
to spatial filtering because it consists of three GMCs and therefore
shows more structure than the other two targets.

Based on the results of the mock observations, we make the following
comments regarding the use of interferometer data alone in measuring
cloud properties. Most of these points can be seen visually in Figure
\ref{interf_props}.
\begin{enumerate}
\item Cloud properties measured from interferometric data are
biased. The degree of bias is affected by the sensitivity of the array
as well as the $(u,v)$ coverage of the observations.
\item A minimum signal-to-noise of 10 is required for stable recovery
of cloud properties.  Below this level, errors in cloud properties,
can approach 100\% for interferometer data.
\item Even for intermediate signal-to-noise values
($T_{peak}/\sigma_{RMS}=10$) there are significant systematic effects
on the cloud properties. The most extreme effects are on the
luminosity measurement, which can be 40\% lower than a single-dish
observation. This effect is much less pronounced for measurements of
the line width and the radius, which show $\lesssim 10\%$
variations. This result, that the radius and line width are relatively
robust to the spatial filtering of the interferometer, confirms the
qualitative results of \citet{she00}. The values of derived properties
are always {\em underestimated} relative to single-dish observations.
\item Even at high sensitivities, the spatial filtering of
interferometers affects property recovery at the 10\% level.  For
example, C-array observations of Orion systematically underestimate
the radius of the cloud by 10\% even for very high signal to noise
ratios and D-array observations underestimate the flux of the Orion by
5\% even at high sensitivity.
\item For interferometer observations, the dynamical mass measurements
of GMCs are more robust than the luminosity measurements.  This
behavior will bias interpretations of the virial parameter in
extragalactic observations.  Estimates of the CO-to-H$_2$ conversion
factor based on the assumption that GMCs are bound or virialized are
likely to {\em overestimate} $X_{\mathrm{CO}}$.
\item For a given signal-to-noise value, observations with the widest
range of $(u,v)$ coverage provide the most robust measurement of cloud
properties.  Thus, in achieving a given sensitivity, observers should
favor arrays with more antennae or observations made in multiple
configurations.
\end{enumerate}

\section{A Note on Decomposition}
\label{DECOMPSECT}

The choice of how to decompose an emission distribution into
individual clouds may be the most important source of bias in GMC
property measurements.  Many different methods have been applied to
identify GMCs in blended emission, the most prevalent being
decomposition by eye and the application of the GAUSSCLUMPS
\citep{gaussclumps} or the CLUMPFIND algorithm \citep{clumpfind}.
When comparing GMCs between two data sets, care must be taken to
decompose emission in a consistent way across both data sets,
preferably using the same algorithm on both data sets. Furthermore,
the physical values of any tuning parameters in the algorithms should
be matched where possible so that both algorithms search for peaks in
the emission over the same spatial scale (rather than angular or
resolution-units) or intensity range. This will avoid, for example,
comparing ``clumps'' in a Galactic molecular cloud to GMCs in another
galaxy. A more extreme method to ensure accurate comparison is to
convolve the higher (spatial) resolution data set to the spatial
resolution of the other data \citep[e.g.~][]{she00}. However, this
approach clearly sacrifices accuracy of the derived parameters to
allow a more careful comparison between two data sets.

In the Appendix to this paper we present a robust, conservative, new
decomposition algorithm.  This algorithm is designed to avoid creating
spurious clouds from noise and to remain sensitive to non-Gaussian
structures in the data. Additionally, the parameters of the algorithm
are fixed to physical values rather than being determined by the
data. This algorithm is designed explicitly with the goal of
decomposing emission into GMCs (rather than clumps or other
structures) with extragalactic data in mind. The methods for
measurement of cloud properties described above --- including the
sensitivity and resolution corrections --- are independent of the
decomposition algorithm and are important no matter what decomposition
algorithm is chosen. In order to avoid confusion between these two
separate problems, we choose to describe the decomposition algorithm
in the Appendix.  

\section{Discussion and Conclusions}
\label{APPLICATIONS}
We conclude the paper by applying these methods to molecular line data
sets that have been previously published. In future studies, the
algorithm will be used to evaluate the differences between GMC
populations between galaxies. Here, we simply present an analysis
designed to demonstrate the method's utility. We present the median
corrections found for a large set of extragalactic (Local Group)
observations and a test application of our methods to Galactic
data. We use all the methods discussed in the previous sections and
the decomposition algorithm discussed in the Appendix. 

\subsection{The Effects of Extrapolation and Deconvolution}

We apply the methods outlined here to an array of data from across the
Local Group and measure the properties of 110 spatially resolved
clouds to estimate the typical magnitude of the sensitivity and
resolution corrections for extragalactic data. We use BIMA data on
M~33 and M~31 \citep{ros03,ros05}; NANTEN observations of the LMC
\citep[][]{fu99}; OVRO observations of IC~10 \citep[][]{wa05}; and a
SEST map of N83 in the SMC \citep[][]{bol03}. Table \ref{corrections}
shows the number of clouds measured in each galaxy along with the
median sensitivity and resolution corrections applied to the radius,
line width, and flux. For comparison, we also measure the properties
of a number of clouds in the outer Galaxy (Quadrant 2, see below) from
the \citet[][]{mwco} CO survey of the Milky Way. Table
\ref{corrections} includes all spatially resolved clouds with
masses (derived from the CO luminosity) of $5 \times 10^4$ M$_{\odot}$
or more ($92$ of the $110$ extragalactic clouds are above this mass).

The numbers quoted in Table \ref{corrections} are ``correction
factors,''

\begin{equation}
\frac{R_{corrected}}{R_{uncorrected}},
~\frac{\sigma_{v,corrected}}{\sigma_{v,uncorrected}},\mbox{ and } 
\frac{L_{\mathrm{CO},corrected}}{L_{\mathrm{CO},uncorrected}} \mbox{ .}
\end{equation}
Table \ref{corrections} shows that throughout the Local Group data the
corrections suggested in this paper have magnitudes of a few tens of
percent. We draw several conclusions based on these data:

\begin{enumerate}
\item Resolution effects on the size of clouds tend to be significant
  --- we would overestimate cloud sizes by $\sim 40$\%\ if we did not
  apply a deconvolution. In the Milky Way data, this effect is much
  less severe. The sizes of Milky Way clouds are measured to within
  $5\%$ before the resolution correction. Unresolved clouds do not
  contribute to Table \ref{corrections}, so if the effects of the
  resolution bias were completely neglected this would be much larger
  (a naive approach would measure these clouds to have the size of the
  spatial beam).
 
\item Resolution effects on the line width are negligible throughout
the Local Group data.

\item Sensitivity effects are also significant. Without a correction
for the sensitivity bias, the size, line width, and luminosity of
clouds would all be significantly underestimated. This sensitivity
bias is least severe --- only about 20 -- 30\%\ --- for the line
width, and most significant (and variable) for the
luminosity. Sensitivity corrections to the luminosity vary from 20\%
to more than 100\%.

\item The resolution and sensitivity biases for the size measurement
tend to cancel out, so that the completely uncorrected radius
measurement is often within 10 -- 20\%\ of the corrected value. This
is a happy coincidence of resolution and sensitivity within the Local
Group, not evidence that sensitivity and resolution corrections are
unimportant.

\item The magnitude of corrections across the Local Group data are
fairly uniform. This is because GMCs near the resolution limit tend to
outnumber higher mass GMCs. Unresolved GMCs are not included in the
analysis, so the median cloud through all the data sets appears
marginally resolved.

\item In order to compare extragalactic data to Galactic data (with
very good sensitivity and resolution and therefore small corrections)
it is crucial to correct for the sensitivity and resolution biases.
\end{enumerate}

\begin{center}
\begin{deluxetable*}{l c c c c c c}

\tablecaption{\label{corrections} Typical Corrections for Local Group Data}

\tablehead{ \multicolumn{1}{l}{Galaxy} & \multicolumn{1}{c}{$N_{Clouds}$} &
\multicolumn{3}{c}{Sensitivity Correction} &
\multicolumn{2}{c}{Resolution Correction} \\
\multicolumn{1}{l}{} & \multicolumn{1}{c}{} &
\multicolumn{3}{c}{$\overbrace{\phm{SpanningSpann}}$} &
\multicolumn{2}{c}{$\overbrace{\phm{SpanningSpann}}$} \\
\multicolumn{1}{l}{} & \multicolumn{1}{c}{} &
\multicolumn{1}{c}{$\frac{R_{corrected}}{R_{uncorrected}}$} & 
\multicolumn{1}{c}{$\frac{\sigma_{v,corrected}}{\sigma_{v,uncorrected}}$} &
\multicolumn{1}{c}{$\frac{L_{CO,corrected}}{L_{CO,uncorrected}}$} & 
\multicolumn{1}{c}{$\frac{R_{corrected}}{R_{uncorrected}}$} & 
\multicolumn{1}{c}{$\frac{\sigma_{v,corrected}}{\sigma_{v,uncorrected}}$} \\}

\startdata
LMC & 46 & $1.4$ & $1.2$ & $1.7$ & $0.8$ & $1.0$ \\
M~31 & 28 & $1.5$ & $1.3$ & $1.6$ & $0.7$ & $1.0$ \\
IC~10 & 17 & $1.7$ & $1.3$ & $2.3$ & $0.7$ & $1.0$ \\
M~33 & 15 & $1.4$ & $1.2$ & $1.5$ & $0.7$ & $1.0$ \\
SMC & 4 & $1.1$ & $1.2$ & $1.2$ & $0.7$ & $1.0$ \\
MW\tablenotemark{a} & 107 & $1.1$ & $1.1$ & $1.4$ & $1.0$ & $1.0$ \\
\enddata
\tablenotetext{a}{Quadrant 2 clouds with $M > 5 \times 10^4$ M$_{\odot}$.}
\end{deluxetable*}
\end{center}

\subsection{Analysis of Second Quadrant CO Data}

The method described in this paper has been designed with
extragalactic data in mind. However, a crucial step in interpreting
extragalactic measurements is to make a fair comparison with Galactic
data. In this section we report some results of applying our
decomposition and measurement algorithms to the survey of the second
Galactic quadrant by \citet{mwco}. We compare the results of this
analysis to the results by \citet[][]{hc01} and show that our analysis
recovers results that are consistent with theirs.

We decompose and analyze $^{12}$CO($1\to 0$) from the second quadrant
\citep[Survey 17 in Table 1 of][]{mwco}. The data set covers the
Galactic plane from $\ell = 70^{\circ}$ to $\ell = 210^{\circ}$ with a
noise level of $0.3$ K. We measure the distance to the molecular
emission using the kinematic distances by adopting a flat rotation
curve with $\Theta_{\mathrm{LSR}} = 220$~km~s$^{-1}$ and
$R_\odot=8.5$~kpc. We omit local emission by discarding all elements
of the data cube with a kinematic distance less than 2 kpc as well as
all elements in the data cube that are connected by significant
emission in position or velocity space to such pixels. We apply the
decomposition algorithm described in the Appendix and measured sizes,
line widths, and luminosities of GMCs using the methods of \S
\ref{EXTRAPSECT} and \S \ref{BEAMDCSECT}. The analysis
recovers 431 clouds with resolved angular sizes and line widths
located within 10 kpc of the Sun. We include the
median sensitivity and resolution corrections for massive ($>5 \times
10^4$ M$_{\odot}$) clouds in Table \ref{corrections} above.

\begin{figure}
\begin{center}
\plotone{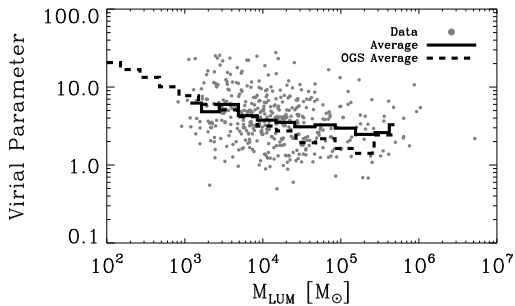}
\caption{\label{virparams} Virial Parameter as a Function of Luminous
Mass for Clouds in the Perseus Arm. Gray dots indicate the 431 clouds
for which the virial and luminous mass can both be determined.  The
solid black line is the average of the data in bins of 0.25 dex in
mass.  For comparison, the average virial parameter is plotted as a
function of mass for the clouds in the OGS catalog (dashed line).  The
two data sets show similar virial parameters at all masses, recovering
nearly the same virial parameter for a given mass.}
\end{center}
\end{figure}

Do the results from our algorithm agree with previous studies of
Galactic GMCs? The data set covers the region studied by \citet{hc01}
using the $45''$ resolution of the FCRAO 14m. That data set has a
lower sensitivity than the \citet[][]{mwco} data, so we apply a
rudimentary sensitivity correction (assuming that the GMCs are
Gaussian and using their peak temperature and boundary isosurface) to
their results and scale to $X_{\mathrm{CO}}=2\times 10^{20}\mbox{
cm}^{-2} (\mbox{K km s}^{-1})^{-1}$ before comparing GMC
properties. We focus on a comparison of the virial parameter between
the two studies --- a full treatment of the Galactic ``Larson's Laws''
is beyond the scope of this paper. We draw several conclusions from
the comparison:

\begin{enumerate}

\item Figure \ref{virparams} shows that the virial parameters derived
in our analysis largely agree with those found by
\citet[][]{hc01}. Below masses of $\approx 3 \times 10^4$ M$_{\odot}$,
both surveys find the same virial parameter in a given mass bin.

\item Above $\approx 3 \times 10^4$ M$_{\odot}$, our analysis of the
\citet[][]{mwco} data set may yield a slightly higher virial
parameter, on average. This may be evidence for the diffuse emission
mentioned in \S\ref{EXTRAPSECT} --- the higher sensitivity
\citet[][]{mwco} data may include diffuse emission surrounding the
GMCs while the \citet[][]{hc01} data may miss this effect. It may also
reflect inadequacies in the simple sensitivity correction we apply to
the \citet[][]{hc01} data. Resolution effects may also play a role ---
the \citet[][]{hc01} data set has $\sim 5$ times better resolution
than the \citet[][]{mwco} data and so the lower resolution data may
tend to lump unbound clouds together. The number of clouds in the high
mass bins is relatively low, so the discrepancy may not be
particularly significant.

\item We apply our algorithm to a small portion of the OGS data and
find our corrections for resolution and sensitivity bias increase the
mean virial parameter to be consistent with the results from the
\citet{mwco} data. This suggests that the differences in the virial
parameters for the high mass clouds in Figure \ref{virparams} may
simply be methodological.

\item Both catalogs of outer Galaxy clouds find an inverse
relationship between luminous mass and the virial parameter,
approximately $\alpha \propto M_{\mathrm{Lum}}^{-0.2}$ --- a
relationship that was also observed by \citet{srby87} for inner Galaxy
molecular clouds.
\end{enumerate}

Thus we find agreement with the results of previous studies of
Galactic molecular clouds. Our methods applied to Galactic data find
the same behavior observed in earlier work and we find agreement among
the method applied to several data sets.

\subsection{Conclusions}

We have presented a method for measuring macroscopic GMC properties
--- spatial size, line width, and luminosity --- from a region of
emission in a spectral line data cube. This method corrects for biases
from limited sensitivity and resolution and produces reliable results
that are directly comparable among a wide variety of data sets. We
correct for limited sensitivity via an extrapolation to a theoretical
0 Kelvin isosurface. We apply a simple quadratic deconvolution to
the extrapolated values to account for resolution biases. We find that
bootstrapping methods yield believable estimates of the uncertainties
in the derived parameters. We present a set of suggestions for
transforming the derived properties into physical quantities of
interest. In the Appendix to this paper we present a new method for
decomposing emission into individual GMCs. This method is conservative
and robust, designed to produce robust results from low
resolution/sensitivity data. In this section we have applied all of
these methods to an array of extragalactic (Local Group) and Galactic
data. We find that the algorithm reproduces established results for
Galactic GMCs and that the sensitivity and resolution biases are
potentially significant --- often $\sim 40\%$ --- for even the most
recent Local Group GMC measurements.

Based on this investigation of the observational biases in measuring
molecular cloud properties, we note several important points that
should be considered in planning observations of GMCs and interpreting
the results.  First, resolution and sensitivity biases can be
corrected to $<10\%$ error provided the cloud has a modest peak
signal-to-noise ($T_{peak}/\sigma_{RMS}\gtrsim 10$) and is marginally
resolved $R_{cld} > 0.8\theta_{\mathrm{FWHM}}$, where
$\theta_{\mathrm{FWHM}}$ is the full width at half maximum extent of
the beam.  Given current and future telescope capabilities, even the
properties of extragalactic GMCs can be accurately measured.  From
Figure \ref{DISTCOMP}, we can see that single dish surveys can
accurately study clouds more massive than $10^4~M_{\odot}$ in the
Magellanic clouds and careful interferometer observations can recover
cloud properties for clouds with mass $M\gtrsim 10^5~M_{\odot}$ in M31
and M33.  However, interferometer observations systematically
underestimate molecular cloud properties.  All else being equal,
interferometers can underestimate fluxes by 40\% and cloud radii by
10\% relative to single-dish observations.  Line widths are largely
unaffected by interferometer observations. The magnitude of the
systematic bias depends on the array configuration and the sensitivity
of the observations.  In general, wider coverage of the $(u,v)$ plane
produces better property recovery.  To minimize bias, observers should
favor observations from several array configurations or from arrays
with many elements.  If possible, the interferometer data should be
supplemented with single-dish observations.  Finally, the
decomposition algorithm used to separate emission into physically
relevant structures will systematically affect molecular cloud
properties.  To date, there is no algorithm that should be favored in
all circumstances, but any comparative study of GMC properties should
be consistent in the choice of algorithm.  For example, the results of
a CLUMPFIND algorithm applied to an extragalactic data set are not
directly comparable to a catalog produced by a simple contouring
method applied to Milky Way data.  Provided the same algorithm is used
across multiple data sets, referencing algorithm parameters to a
common physical scale will minimize systematic differences.

We will make a software version of the decomposition and measurement
algorithm available as an IDL package. Further, since methodology can
affect the results of GMC studies so strongly, we encourage authors
working in the field to make their data available to the community
after publication in order to facilitate future rigorous comparisons.

\acknowledgements

We are extremely grateful to Tom Dame and the Millimeter-Wave Group at
the Center for Astrophysics for providing both the Orion data and the
Quadrant 2 data used in the paper. We also thank the NANTEN Group at
Nagoya, especially Yasuo Fukui and Akiko Kawamura, providing us with
the LMC CO data. We are grateful to Fabian Walter for providing us
with the OVRO IC~10 data. Alberto Bolatto, Jason Wright, Jon Swift,
and Leo Blitz all offered helpful comments on drafts of the paper.  We
thank Ronak Shah for helping us compare our IDL version of GAUSSCLUMPS
to the original implementation of the algorithm.  The informed
comments of an anonymous referee greatly improved the paper,
particularly in encouraging us to explore the effects of
interferometers.  ER is grateful for support through the National
Science Foundation Astronomy \& Astrophysics Postdoctoral Fellows
Program (AST-0502605). This work is partially supported by NSF grant
0228963 to the Radio Astronomy Laboratory at UC Berkeley.

\begin{appendix}

\label{ouralg}
\section{Appendix: A New Decomposition Algorithm \label{DECOMPOSITION}}

The choice of what emission to identify as a GMC may be the single
largest source of uncertainty in measuring and comparing GMC
properties. A number of methods have been employed over the years,
from simple contouring methods
\citep[e.g.~][]{sss85,dect86,srby87,hc01} to fitting three-dimensional
Gaussians \citep[GAUSSCLUMPS,][]{gaussclumps,gaussclumps2}, to
modified watershed algorithms \citep[CLUMPFIND and its
kin,][]{clumpfind,clumpfind-mod}.

In this section we present a new decomposition algorithm that is
designed to identify clouds at low sensitivities while avoiding the
introduction of a false clouds due to noise. This algorithm consists
of two parts: identifying regions of significant emission in the data
set and then assigning this emission to individual ``clouds.''

\subsection{Signal Identification}

We first identify regions of contiguous, significant emission in our
position-position-velocity data cube. We estimate the noise in the
data set by measuring the RMS intensity, $\sigma_{RMS}$, from a
signal-free region of the data cube. We then construct a
high-significance mask. This mask includes only adjacent channels that
both have intensities above $4\sigma_{RMS}$. We expand that mask to
include all emission above a lower threshold --- typically two
channels above $2\sigma_{RMS}$ significance --- that is connected to
the original high significance mask through pixels with $\ge
2\sigma_{RMS}$ significance.  The resulting mask contains most of the
significant emission in the data cube. Lowering the threshold below
two channels at $\approx 1.5\sigma_{RMS}$ runs the risk of biasing the
moment measurements towards high values by including false emission
(noise with positive values) in the cloud.

\subsection{Cloud Identification}
\label{algorithm}
In this section, we describe the algorithm used to decompose a region
of emission into individual subsections representing the physically
distinct entities in the data (``clouds'').  Through the description
of the algorithm, there are several parameters that can be varied to
produce changes in the resulting decomposition.  In general, we set
these parameters using physical prior knowledge of the GMCs we are
seeking to catalog.  We discuss the choice of these parameters in the
following section.

\begin{enumerate}
\item {\em Discard Small or Low Contrast Regions:} If a region is too
small for us to measure meaningful properties from it, we discard the
region. We require that each region has an area larger than two beam
sizes, so that we can measure its size; and a velocity width of more
than a single channel, so that we can measure its line width. If the
intensity contrast between the peak and the edge of the region is less
than a factor of two, we lack the dynamic range needed to correct the
sensitivity bias and we therefore discard the region. If a region is
not discarded we proceed to the next step.

\item \label{xform}{\em Rescale the Data to Reduce the Effects of
Substructure:} Molecular clouds contain significant substructure that
confuses the decomposition of these sources.  The substructure is
often significantly brighter than bulk of the gas in the cloud.  We
rescale the data to reduce the contrast between this substructure and
the cloud as a whole.  The data are rescaled using the following
transform:
\[ T' = \left\{\begin{array}{cl}
T & ;~T<T_{clip}, \\
T_{clip} [1+{\arctan (T/T_{clip}-1)}] & ;~ T\ge T_{clip}. \\
\end{array} \right. 
\]
This transform reduces the contrast pixels with $T\gtrsim 2 T_{clip}$
while preserving the relative brightness distribution.  The value of
$T_{clip}$ is left as a free parameter.  The transformed data are used
in the decomposition algorithm.  Such brightness transforms are
frequently used in the decomposition algorithms used in other fields
such as medical imaging \citep[e.g.~][]{cytom}.

\item {\em Identify Independent Local Maxima:} We identify potential
local maxima, by identifying the elements in the data cube that are
larger than all their neighbors. We consider neighbors to be all data
that lie within a box with side length $D_{max}$ in position and
$\Delta V_{max}$ in velocity centered on the local maximum.  These are
our ``candidate maxima.''  The parameters $D_{max}$ and $\Delta
V_{max}$ are free parameters.  If we find more than one candidate
maximum in a region of emission, we proceed to the next steps to
further verify each maximum's independence. If we find a single
candidate maximum then we label the region as a cloud and measure its
properties as described in the main paper.\label{locmax}

\item {\em Find Shared Isosurfaces and Reject Small Clouds or those
with Smooth Mergers:} For each {\em pair} of candidate maxima we
calculate the value of the highest intensity isosurface to contain
both maxima. We refer to this highest shared isosurface as the {\em
merge level}.  Using this set of highest shared isosurfaces, we
calculate three properties of interest:
\begin{enumerate}
\item The area uniquely associated with each maximum (i.e. the area
above the merge level for that maximum).
\item The antenna temperature interval between the merge level and
 each maximum, referred to as the {\em contrast
 interval}. \label{decimate}
\item The fractional amount by which each of the (unextrapolated)
moment values changes across each shared isosurface
(i.e. $\frac{\Delta \sigma_{maj}}{\sigma_{maj}}$ for each maximum
across each shared isosurface).  We consider the two clouds to merge
smoothly across the isosurface only if:
\begin{enumerate}
\item None of the second moments increase by more than 100\% for both maxima.
\item No two of the second moments increase by more than 50\% for both
maxima.
\item The flux increases by less 200\% for both maxima.
\end{enumerate}
\end{enumerate}
We use these three properties to pare maxima from the region. We
reject maxima associated with small areas (less than two beam sizes,
as for the region above) and contrast intervals less than $\Delta
T_{max}$, a free parameter.  Choosing $\Delta T_{max}\ge 2\sigma$
significantly reduces the effects of noise on decomposition
\citep[e.g.~][]{clumpfind-mod} since noise is associated with low
contrast intervals.  Finally, when a pair of maxima merge smoothly
across a shared isosurface, we keep the higher intensity maximum and
discard the lower one.  This is a conservative choice in the
decomposition algorithm: unless merging the two kernels significantly
alters the properties of one of the clouds associated with the
separate kernels, we assume the kernels are not physically distinct.
The effect of removing kernels that merge smoothly from our data set
is to reduce the algorithm's sensitivity to substructure within
clouds.  We iterate this step until we have a set of maxima associated
with the required areas and separated from each other by significant
jumps in their properties.

\item {\em Define Clouds Using Shared Isosurfaces:} The surviving
maxima each correspond to a ``cloud.'' That cloud consists of the
emission within the lowest intensity isosurface uniquely associated
with the cloud. Emission that lies below this isosurface is part of a
``watershed'' shared among clouds and we do not assign it to any
cloud.  By not considering contested emission, i.e.~emission that
could be associated with distinct local maxima, we avoid the problem of
how to properly assign this emission to local maxima.

\item {\em Measure Cloud Properties:} Finally, we apply the methods
described in \S2 and 3 to derive spatial sizes, line widths, and
luminosities for each cloud.  The transformed data (Step
\ref{xform}) are inverse transformed into the original brightness
units for this analysis.
\end{enumerate}

\subsection{Using Physical Priors to Establish Algorithm Parameters}
 
Several of the algorithm's parameters are left to the choice of the
user.  Without any prior knowledge of the physical objects in the
data, we establish default values for these parameters that will
produce a reasonable decomposition based solely on the characteristics
of the data.  These defaults are chosen to provide sensitivity to real
substructure within the data without contamination by noise and can be
regarded as the {\it minimum} appropriate values for these parameters
in most cases.  The free parameters in the algorithm are the
brightness transform threshold ($T_{clip}$, Step \ref{xform}) position
and velocity window size used in searching for local maxima ($D_{max}$
and $\Delta V_{max}$, Step \ref{locmax}), and the minimum contrast
temperature ($\Delta T_{max}$, Step \ref{decimate}).  Without prior
knowledge of substructure in the data, no brightness transform should
be applied ($T_{clip}=\infty$).  The minimum values for the parameters
in the search for initial local maxima are set by the resolution of
the data: $D_{max}=\theta_{beam}$ and $\Delta V_{max}=\Delta V_{chan}$
or else separations between local maxima cannot be resolved.  Finally,
we use $\Delta T_{max}=2\sigma_{RMS}$ to prevent the noise in the data
set from being recognized by the algorithm as legitimate structure and
becoming the basis for the decomposition.

In the main section of the paper, we developed methods that measured
physical properties of molecular emission without observational bias.
Ideally, the decomposition algorithm used in the analysis would also
be free of observational bias.  A good algorithm would, for example,
decompose the same emission into similar structures regardless of the
resolution and sensitivity of the observations.  To progress towards
this goal, we set the parameters of the decomposition algorithm to
have similar values in physical units (pc, km s$^{-1}$, K) rather than
data units (beam widths, channels, $\sigma_{RMS}$).

In the molecular ISM, which has structure on a range of scales, the
choice of the physical values for the algorithm parameters must be
motivated by prior knowledge of the objects that we wish to identify.
The choice of parameters for identifying GMCs would be different from
the choice of parameters for identifying clumps and cores.  As a
population, GMCs have size scales of 10s of parsecs, line widths of
several km s$^{-1}$, and brightness temperatures of $T
\lesssim 10$~K \citep[e.g.~][]{srby87}.  In contrast, the clumpy
substructure within clouds has a size scale of 1 pc, line widths of
order 1 km s$^{-1}$ and can have brightness temperatures of 30 K or
higher \citep[e.g.~][]{clumpfind}.  We select the parameters of the
algorithm to find molecular clouds rather than the clumps within them.
In particular, we set the minimum separation between local maxima as
$D_{max}=15$~pc and the velocity separation to be $\Delta V_{max}
=2$~km s$^{-1}$.  We also fix a contrast interval in antenna
temperature of $\Delta T_{max} = 1$~K rather than the data-driven
value of $2 \sigma_{RMS}$.  Finally, we must account for the presence
of bright molecular gas substructure within molecular clouds.  For
example, Orion A has a few distinct regions separated by $>15$ pc with
antenna temperatures in excess of 15 K \citep{wil05}.  These are
associated with hot molecular gas around young stars like the
Trapezium cluster.  In general, the typical kinetic temperature of gas
in GMCs is $\sim 10$~K so brightness temperatures in excess of 10 K
are usually associated with structure {\it within} molecular clouds
not changes from cloud to cloud.  Hence, to catalog clouds and not
substructure, we must reduce the influence of the high $T_b$ data with
the parameter $T_{clip}$.  We use $T_{clip} = 2.5$~K to maintain the
full sensitivity for the gas separating GMCs (which typically has $T_A
\lesssim 5$~K in the \citet{srby87} data) while reducing the influence
of brighter gas associated with a single GMC.  Using $T_{clip}=2.5$~K
has little influence on extragalactic data where beam deconvolution
typically averages out the presence of bright substructure within the
clouds.

We summarize our choices of parameters in Table \ref{params} when the
parameters are motivated by the data (Data-Based) or by physical
assumptions about the structures being extracted (GMC Physical Priors).  The
resolution or the sensitivity of a data set may be sufficiently poor
that the physical parameters are unattainable in the data set
(e.g.~$\theta_{beam}>15~$pc, $2\sigma_{rms}> 1$~K).  In this case, the
decomposition must be regarded with caution as it may not be directly
comparable to other data sets.  The adopted values are appropriate for
decomposing data sets of $^{12}$CO emission.  The values of $T_{clip}$
and $\Delta T_{max}$ would be different for other tracers.
\begin{deluxetable}{lcc}
\tablecaption{Parameter Choices for Decomposition Algorithm\label{params}}
\tablehead{
\colhead{Parameter} & \colhead{Data-Based} & \colhead{GMC Physical
Priors}}\
\startdata
$T_{clip}$ & $\infty$ & 2.5 K \\
$D_{max}$ & 1 beam width & 15 pc \\
$\Delta V_{max}$ & 1 channel & 2 km s$^{-1}$ \\
$\Delta T_{max}$ & $2\sigma_{RMS}$ & 1 K \\
\enddata
\end{deluxetable}

\subsection{Comparison with Existing Algorithms}

To demonstrate that the proposed algorithm is actually an improvement
on existing methods, we analyze trial data sets with known properties
using this algorithm and compare the results to those from the
GAUSSCLUMPS and CLUMPFIND algorithms to the trial data.  We use the
CLFIND algorithm implemented in MIRIAD \citep{miriad} and our own IDL
implementation of the \citet{gaussclumps2} GAUSSCLUMPS algorithm,
which we have compared to the standard algorithm operating on a real
data set with satisfactory results. 

\begin{figure}
\begin{center}
\plotone{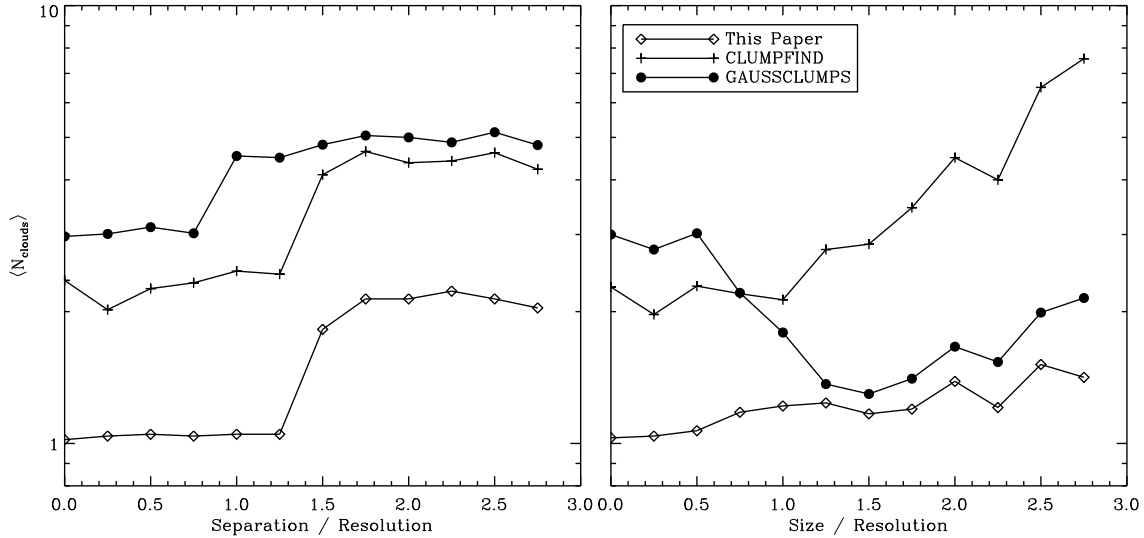}
\figcaption{\label{nclplot} The mean number of clouds with 
$T_{peak}/\sigma_{RMS}>5$ recovered by three different decomposition
algorithms on trial data. (left) The three algorithms applied to a
blend of two Gaussian clouds. The results of using algorithm from the
present study are minimally affected by the presence of noise.  (right) The
three algorithms applied to a circular top-hat cloud showing the
effects of non-Gaussian shape on the results of the algorithms.  The
current algorithm is the best for grouping together large structures
with non-Gaussian shapes.}
\end{center}
\end{figure}

We first examine how adept the three algorithms are at decomposing a
pair of blended clouds.  We construct a series of data cubes with two
unresolved model clouds separated in position by a variable distance.
The model clouds each have a peak signal-to-noise ratio of 10 with a
Gaussian line profile.  For each value of the separation, we decompose
100 data cubes with different realizations of the noise using the
three algorithms, using data based choices for the algorithm
parameters.  Figure \ref{nclplot} (left) shows the mean number of
clouds recovered by each algorithm as a function of the separation (we
count clouds with peak signal-to-noise larger than 5 as
``recovered''). As the trial clouds are moved farther apart, the
typical number of clouds detected by each algorithm increases by
1. However, only the algorithm presented here consistently recovers a
single cloud at low separations. GAUSSCLUMPS and CLUMPFIND produce
false clouds from the noise. The jump in the number of clouds detected
by the GAUSSCLUMPS algorithm occurs where the separation equals the
resolution, whereas both the current algorithm and CLUMPFIND are able
to distinguish the clouds only if their separation is over 1.5 times
the resolution. GAUSSCLUMPS appears to be able to distinguish tight
blends of clouds but is the most susceptible to noise.

As a second test of the algorithms, we compared the number of clouds
that the algorithms recover from a single cloud with a circular
top-hat brightness profile ($T_A=\mbox{const. for } r<R_0;~0$
otherwise) and peak signal-to-noise of 10. The size of the cloud is
varied with respect to the resolution of the data set and 100 data
sets for each value of $R_0$ are decomposed by each algorithm.  The
non-Gaussian brightness profile confounds all of the algorithms but to
varying degrees.  CLUMPFIND detects an increasing number of spurious
clouds as the cloud grows, suggesting that the number of false clouds
grows with the volume studied.  Despite the non-Gaussian profile,
GAUSSCLUMPS does surprisingly well with large sources.  The current
algorithm, however, does the best job of detecting a single source in
the presence of noise.  For analyzing data with a relatively low
signal-to-noise, we find the decomposition algorithm presented in this
Appendix should be favored for identifying clouds.

\subsection{Applying the Algorithm to the Orion-Monoceros Region}
\begin{figure}
\begin{center}
\plotone{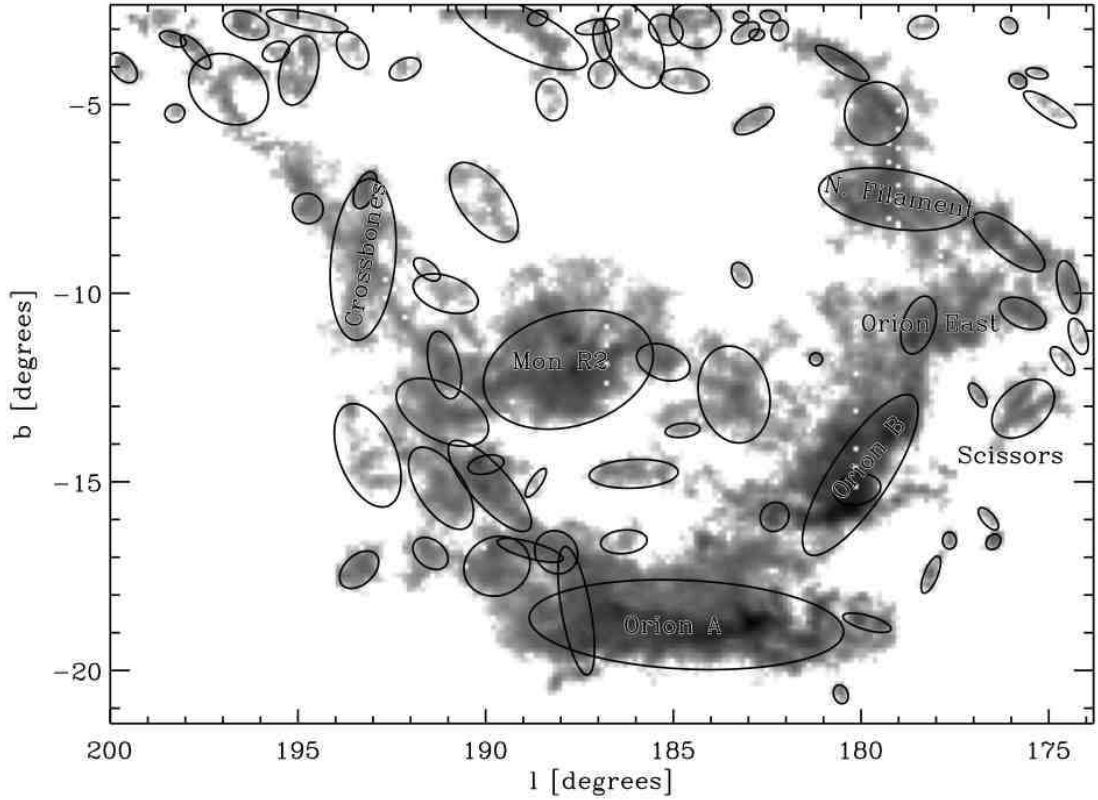}
\figcaption{\label{orion_fig} Masked, integrated-intensity map of the
Orion-Monoceros region with a logarithmic stretch.  The ellipses
plotted over the data indicate the positions, sizes and orientations
of molecular clouds identified by our decomposition algorithm.  In
general, the algorithm does a good job of identifying known molecular
clouds as single entities. Notable molecular features are labeled.}
\end{center}
\end{figure}

The Orion Molecular Cloud is among the best studied of all GMCs, so it
is an good place to compare the results of our methods to those of
previous work.  For this comparison, we use the data and results of
the recent, uniform survey of the entire Orion-Monoceros region by
\citet{wil05}.  We analyze their final data set using the methods in
this paper (with physical priors for the decomposition algorithm
parameters) and summarize the results in Figure \ref{orion_fig}, an
integrated intensity map of the region. The decomposition results in
81 molecular clouds, most of which are associated with the Galactic
plane at the top of the Figure.

The results of the algorithms decomposition of the region are good:
major molecular clouds are identified as single entities in the
decomposition. The algorithm succeeds where other cloud identification
schemes would face difficulty.  Nearly all of the emission in the data
is connected above a single isosurface so that simple contouring
methods would identify the entire Orion-Monoceros region as a single
molecular cloud, though more complicated algorithms \citep{syscw} may
succeed. CLUMPFIND and GAUSSCLUMPS would isolate individual peaks and
decompose clouds into their substructure --- as was intended by their
design (CLUMPFIND identifies 617 clumps in the same data).

The properties of the clouds agree well with the values published for
a human decomposition of the emission.  The results of the analysis
are given in Table \ref{orion_props}. For comparison, we adopt the
distances of \citet{wil05}.  We compare the results of the algorithm
to their results, after scaling their mass up by 10\% to account for a
difference in the adopted CO-to-H$_2$ conversion factors.  In several
cases the masses agree quite well (to within 5\%, Orion A, Mon R2,
Scissors), while other features show $\sim 30\%$ differences.  These
systematic discrepancies arise from differences in how emission is
assigned into structures.  For example, the algorithm only identifies
the central region of Orion B as a molecular cloud; it does not
include emission near the location of Orion East that is nominally
part of the Orion B cloud because the assignment of this emission is
contested with neighboring clouds.  Similarly, the algorithm
characterizes the Northern Filament region as five distinct clouds
rather than the single large cloud that a human decomposition
produced.  Despite these differences, the results of the algorithm are
reassuring -- in most cases, the well-known molecular clouds are
identified as single clouds and there is good agreement between their
derived physical properties and the results of previous studies.

\begin{deluxetable}{lcccccc}
\tablecaption{\label{orion_props} Properties of Major Molecular Clouds in 
Orion-Monoceros}
\tablewidth{0pt}
\tablehead{
\colhead{Name} & \colhead{Distance}\tablenotemark{a} 
& \colhead{$M_{\mathrm{human}}$}
\tablenotemark{a} & 
\colhead{$M_{\mathrm{algorithm}}$} & \colhead{$R_e$} & \colhead{$\sigma_v$} &
\colhead{$\alpha$} \\
& \colhead{(pc)} & \colhead{($10^4 M_\odot$)} & 
\colhead{($10^4 M_\odot$)} & \colhead{(pc)} & \colhead{(km s$^{-1}$)} & 
}
\startdata
Orion A & 480 & 12. & $11. \pm 0.06$ & $18.6 \pm 0.2$ & $2.9 \pm 0.1$ & $1.6 \pm
 0.1$ \\
Orion B & 500 & 9.1 & $5.6 \pm 0.08$ & $12.1 \pm 0.3$ & $1.5 \pm 0.1$ & $0.6 \pm
 0.1$ \\
Orion East & 120 & 0.013 & $0.013 \pm 0.001$ & $1.2 \pm 0.1$ & \nodata
& \nodata \\
Mon R2 & 800 & 12. & $11. \pm 0.5$ & $25.9 \pm 1.2$ & $1.6 \pm 0.1$ & $0.7 \pm 0
.1$ \\
Crossbones & 470 & 1.9 & $0.57 \pm 0.02$ & $11.0 \pm 0.6$ & $0.8 \pm 0.1$ & $1.5
 \pm 0.3$ \\
Northern Filament & 390 & 1.9 & $1.2 \pm 0.07$ & $8.6 \pm 0.7$ & $1.7 \pm 0.1$ &
 $2.6 \pm 0.5$ \\
\enddata
\tablenotetext{a}{Adapted from Table 2 of \citet{wil05}.}
\end{deluxetable}

\subsection{Decomposing Orion-like Clouds in Other Galaxies}
\begin{figure}
\begin{center}
\plotone{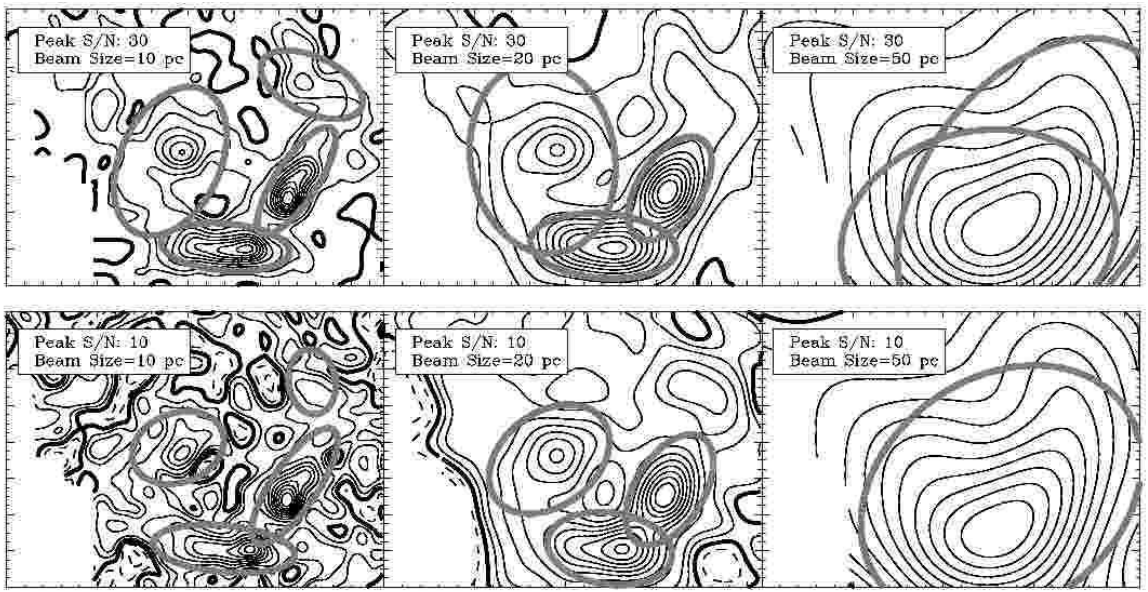}
\figcaption{\label{decompdemo} The Orion-Monoceros Region convolved to
the typical signal-to-noise ($S/N$) and spatial resolution (beam size)
of extragalactic observations.  The integrated intensity map of the
cloud is drawn as a contour plot with the positions and sizes of
clouds that the algorithm identifies drawn as gray ellipses.  Contours
on the integrated intensity map are drawn at -0.1 (dashed), 0 (bold),
0.1, 0.2, ... 1.0 times the peak value in the map.  The sizes of the
clouds are {\it not} deconvolved from the beam, so coarser spatial
resolution results in objects that are apparently larger.  Except in
the marginally resolved, low $S/N$ case, the algorithm successfully
identifies Orion A and B.  Mon R2 and the Northern Filament appear in
maps with higher resolution and signal-to-noise.}
\end{center}
\end{figure}

As a final test of the decomposition and analysis algorithm, we apply
the algorithm to simulated observations of the Orion-Monoceros region
with beam sizes and signal-to-noise values typical of extragalactic
GMC observations.  We simulate observations by convolving the data set
to the desired resolution and adding a convolved data cube of noise
which is scaled up to give the desired peak signal-to-noise ratio in
the data.  We perform this for a peak signal-to-noise ($S/N$) values
of 10 and 30 combined with beam sizes of 10, 20 and 50 pc.  The
results of the decompositions are displayed in Figure
\ref{decompdemo}.

At coarse resolution and low sensitivity, the fainter clouds are
undetectable.  However, the algorithm identifies each of Orion A,
Orion B, Mon R2 and the Northern Filament in at least one of the trial
data sets.  At high resolution (10 pc) and peak signal-to-noise
($S/N=30$), the algorithm successfully identifies the clouds with
properties that are consistent, within the uncertainties determined by
the algorithm, with the masses of the clouds in Table
\ref{orion_props}.  The systematic effects of poor resolution manifest
themselves for beam sizes $\gtrsim 20$ pc where the decomposition of
blended emission results in $\sim 20\%$ variations in the properties
of the clouds relative to those found in the original data set for
both high and low $S/N$.  For a beam size of 20 pc, the clouds are not
resolved since this is over twice the size of the clouds along their
minor axes.  Finally, at 50 pc resolution, Mon R2 is not found by the
algorithm and high $S/N$ is required to distinguish Orion A and Orion
B.  There is a variation of 100\% (0.3 dex) in the derived physical
parameters for the clouds; this variation is reflected in the
estimates of the uncertainties.  As expected based on the 20 pc
resolution data, the clouds are not spatially resolved for a beam size
of 50 pc.  An accurate decomposition of emission into typical Galactic
GMCs requires a beam size $\sim 20$ pc though only a modest
sensitivity is required: $S/N\gtrsim 10$.

\end{appendix}

\end{document}